\documentclass[preprintnumbers,twocolumn,prd,aps,superscriptaddress,footinbib,amsfonts,amsmath,amssymb,showpacs,floatfix]{revtex4-1}

\usepackage[T1]{fontenc}
\usepackage{blindtext}

\usepackage{graphics,graphicx}
\usepackage{dcolumn}
\usepackage{bm}
\usepackage{epsfig}
\usepackage[usenames]{color}
\usepackage[hidelinks]{hyperref} 
\usepackage{mathbbol}
\usepackage{epstopdf}
\usepackage{simplewick}
\usepackage[utf8]{inputenc} 
\usepackage{slashed}
\usepackage{soul}
\usepackage[dvipsnames]{xcolor}
\usepackage[export]{adjustbox}

\def\kT{k_T}
\def\pperp{p_\perp}
\def\Pperp{P_\perp}

\def\avkT{\la \kT^2 \ra}
\def\avpperp{\la \Pperp^2 \ra}

\newcommand{\la}{\langle}
\newcommand{\ra}{\rangle}

\begin{document}

\preprint{JLAB-THY-20-3151}

\title{\bf Origin of single transverse-spin asymmetries in high-energy collisions}

\newcommand*{\WM}{Physics Department, William \& Mary, Williamsburg, Virginia 23187, USA}\affiliation{\WM} 
\newcommand*{\LVC}{Department of Physics, Lebanon Valley College, Annville, Pennsylvania 17003, USA}\affiliation{\LVC}
\newcommand*{\PSU}{Division of Science, Penn State University Berks, Reading, Pennsylvania 19610, USA}\affiliation{\PSU}
\newcommand*{\UCLA}{Department of Physics and Astronomy, University of California, Los Angeles, California 90095, USA}\affiliation{\UCLA}
\newcommand*{\MB}{Mani L.~Bhaumik Institute for Theoretical Physics, University of California, Los Angeles, California 90095, USA}\affiliation{\MB}
\newcommand*{\CF}{Center for Frontiers in Nuclear Science, Stony Brook University, Stony Brook, New York 11794, USA}\affiliation{\CF}
\newcommand*{\JLAB}{Thomas Jefferson National Accelerator Facility, Newport News, VA 23606, USA}\affiliation{\JLAB}
\newcommand*{\ODU}{Department of Physics, Old Dominion University, Norfolk, Virginia 23529, USA \\
            \vspace*{0.2cm}
            {\bf Jefferson Lab Angular Momentum (JAM) Collaboration}}\affiliation{\ODU}

\author{Justin Cammarota}\email{jcammarota@email.wm.edu}\affiliation{\WM}\affiliation{\LVC}
\author{Leonard Gamberg}\email{lpg10@psu.edu}\affiliation{\PSU}
\author{Zhong-Bo Kang}\email{zkang@physics.ucla.edu}\affiliation{\UCLA}\affiliation{\MB}\affiliation{\CF}
\author{Joshua~A.~Miller}\email{jam017@lvc.edu}\affiliation{\LVC}
\author{Daniel Pitonyak}\email{pitonyak@lvc.edu}\affiliation{\LVC}
\author{Alexei Prokudin}\email{prokudin@jlab.org}\affiliation{\PSU}\affiliation{\JLAB}
\author{Ted C. Rogers}\email{tcrogers@jlab.org}\affiliation{\JLAB}\affiliation{\ODU}
\author{Nobuo Sato}\email{nsato@jlab.org}\affiliation{\JLAB}

\begin{abstract}
\noindent 
In this paper we perform the first simultaneous QCD global analysis of data from  semi-inclusive deep inelastic scattering,
Drell-Yan, $e^+e^-$ annihilation into hadron pairs, and proton-proton
collisions. Consequently, we are able to extract a universal set of
non-perturbative functions that describes the observed asymmetries in
these reactions. The outcome of our  analysis indicates single transverse-spin asymmetries  in
high-energy collisions have a common origin. Furthermore, we achieve the first phenomenological
agreement with lattice QCD on the up and down quark tensor charges.
\end{abstract}

\pacs{}
\maketitle

\section{Introduction}
For some fifty years, the spin and momentum structure of hadrons has
been investigated in  terms of their partonic (quark and gluon)
content within the theory of Quantum Chromodynamics (QCD).  
Single transverse-spin asymmetries (SSAs) have played a central role
in these studies. 
Early predictions from QCD that SSAs in single-inclusive hadron
production should be exceedingly small~\cite{Kane:1978nd} were in
stark contrast with measurements showing large
asymmetries~\cite{Bunce:1976yb,Klem:1976ui} that  persist in recent
experiments~\cite{Adams:1991rw, 
      Krueger:1998hz, 
      Allgower:2002qi, 
      Adams:2003fx,
      Adler:2005in, 
      Lee:2007zzh, 
      Abelev:2008af, 
      Arsene:2008aa,
      Adamczyk:2012qj, 
      Adamczyk:2012xd, 
      Bland:2013pkt, 
      Adare:2013ekj,
      Adare:2014qzo, 
      Airapetian:2013bim, 
      Allada:2013nsw}.

A better understanding of SSAs has emerged
with the aid of QCD factorization theorems
\cite{Qiu:1991pp,Qiu:1991wg,Collins:1981uk,Collins:1984kg,Meng:1995yn}.
They separate cross sections into
perturbatively
calculable scattering contributions and 
non-perturbative physics 
encoded in parton
distribution functions (PDFs) and fragmentation functions (FFs).

For processes with one large measured scale, $Q\gg \Lambda_{\rm QCD}$,
where $\Lambda_{\rm QCD}$ is a typical hadronic mass, 
experiments are sensitive to the collinear motion of partons.  For example, in $p^\uparrow p\to h\,X$, the hard scale is set by the hadron transverse momentum $P_{hT}$.
In this case, collinear twist-3  (CT3)
factorization~\cite{Qiu:1991pp,Qiu:1991wg} is valid, and spin
asymmetries arise due to the quantum mechanical interference from
multi-parton states~\cite{Efremov:1981sh,
      Efremov:1984ip,
      Qiu:1991pp,
      Qiu:1991wg,
      Qiu:1998ia,
      Eguchi:2006qz,
      Kouvaris:2006zy,
      Eguchi:2006mc,Koike:2009ge, Kang:2011hk, Metz:2012ct, Beppu:2013uda}.

For reactions with two scales $Q_2\gg Q_1\sim
\Lambda_{\rm QCD}$, experiments probe also
intrinsic transverse parton motion. For example, in semi-inclusive lepton-nucleon deep inelastic scattering
(SIDIS), $\ell\,N \to \ell\,h\,X$, one has $\Lambda_{\rm QCD}\sim P_{hT}\ll Q$, where $-Q^2$ is
the photon virtuality. For such processes, transverse momentum
dependent (TMD) factorization~\cite{Collins:1981uk,
      Collins:1984kg,
      Meng:1995yn,
      Ji:2004xq,
      Collins:2011zzd}
is valid, and the mechanism responsible for spin asymmetries is
encoded in TMD PDFs and FFs (collectively called
TMDs)~\cite{Kotzinian:1994dv,
      Mulders:1995dh,
      Boer:1997mf,
      Bacchetta:2006tn, 
      Arnold:2008kf, 
      Pitonyak:2013dsu}.
      
There are theoretical calculations that use CT3 and TMD factorization theorems to
yield a unified picture of spin asymmetries in hard
processes~\cite{Ji:2006ub,
      Ji:2006br,
      Koike:2007dg,
      Zhou:2008fb,
      Yuan:2009dw,
      Zhou:2009jm}.
This is one of the cornerstones  for studying the 3-dimensional
structure of hadrons at existing~\cite{Aschenauer:2015eha,
    Gautheron:2010wva,
    Bradamante:2018ick,
    Dudek:2012vr,
    Kou:2018nap} 
and future facilities, including  the Electron-Ion
Collider~\cite{Boer:2011fh,Accardi:2012qut}. 
In this paper, we provide, for the first time, phenomenological results that indicate SSAs have a common origin. We perform the first
simultaneous QCD global analysis of the available data in SIDIS, Drell-Yan
(DY), semi-inclusive $e^+e^-$ annihilation (SIA), and proton-proton
collisions.  Furthermore, we find, for the first time, excellent
agreement with lattice QCD for the up and down quark tensor charges. 

\section{Theoretical Background}
The key observation that makes our analysis  possible is that
in both the CT3 and TMD formalisms, collinear multi-parton
correlations play an important role. A generic TMD PDF $F(x,k_T)$ depends on $x$, the
fraction of the nucleon's longitudinal momentum carried by
the parton, and $k_T\equiv |\vec{k}_T|$, the parton's transverse momentum.
The same TMD when 
Fourier conjugated into position ($b_T$) space 
~\cite{Collins:1981uw,Boer:2011xd,Aybat:2011ge,Collins:2011zzd}
exhibits an Operator Product Expansion (OPE) 
in the limit when $b_T$ is small. TMDs relevant for SSAs can be expressed in terms of CT3 multi-parton correlation functions in this OPE~\cite{Aybat:2011ge,
       Kanazawa:2015ajw,
       Gamberg:2017jha,
       Scimemi:2019gge}.

Another way to establish the connection between 
CT3 functions and TMDs is by the use of parton model identities. 
One such relation, derived at the level of QCD-operators, is~\cite{Boer:2003cm}
\begin{equation}
\pi F_{FT}(x,x)=\int \! d^2 \vec{k}_T\,\frac{k_T^2} {2M^2} f_{1T}^\perp(x,k_T^2)
                \equiv f_{1T}^{\perp(1)}(x) \, ,
\label{eq:relation}
\end{equation}
where $F_{FT}(x,x)$ is the Qiu-Sterman CT3 matrix
element, and $f_{1T}^{\perp(1)}(x)$ is the first moment of the TMD
Sivers function $f_{1T}^\perp(x,k_T^2)$ \cite{Sivers:1989cc,Sivers:1990fh}. 
Here we do not address the validity of this relation beyond leading order~\cite{Aybat:2011ge,Kanazawa:2015ajw,Gamberg:2017jha,Scimemi:2019gge,Qiu:2020oqr}.

A central focus of TMD asymmetries 
has been on the Sivers and Collins SSAs in SIDIS, 
$A_{UT}^{\sin(\phi_h-\phi_S)}\!\equiv\! A_{\rm SIDIS}^{\rm Siv}$~\cite{Airapetian:2009ae, Alekseev:2008aa,Qian:2011py,Adolph:2014zba,Zhao:2014qvx,Adolph:2016dvl}
and 
$A_{UT}^{\sin(\phi_h+\phi_S)}\!\equiv\! A_{\rm SIDIS}^{\rm Col}$~\cite{Airapetian:2010ds, Alekseev:2008aa,Qian:2011py,Adolph:2014zba,Zhao:2014qvx}; 
Sivers SSA in DY,
$A_{\rm DY}^{\rm Siv}$, for $W^\pm\!/Z$ production $\!\equiv\! A_N^{W/Z}$~\cite{Adamczyk:2015gyk} and for
$\mu^+\mu^-$ production $\!\equiv\! A_{T,\mu^+\mu^-}^{\sin\phi_S}$~\cite{Aghasyan:2017jop}; 
and Collins SSA in SIA, 
$A_{\rm SIA}^{\rm Col}$
~\cite{Seidl:2008xc,TheBABAR:2013yha,Aubert:2015hha,Ablikim:2015pta,Li:2019iyt}.  
The relevant TMDs probed by these processes~\cite{Kotzinian:1994dv, Mulders:1995dh, 
      Boer:1997mf,
      Bacchetta:2006tn, 
      Arnold:2008kf, 
      Pitonyak:2013dsu} are the
transversity TMD     $h_1(x,k_T^2)$~\cite{Ralston:1979ys}, 
the Sivers function  $f_{1T}^\perp(x,k_T^2)$~\cite{Sivers:1989cc,Sivers:1990fh}, 
and Collins function $H_{1}^\perp(z,z^2p_\perp^2)$~\cite{Collins:1992kk}. 
Each of them can be written in a model-independent way in terms of a collinear counterpart using
the OPE. 
The function $h_1(x,k_T^2)$ is related to the collinear (twist-2)
transversity function $h_1(x)$~\cite{Bacchetta:2013pqa}; 
$f_{1T}^\perp(x,k_T^2)$ to the Qiu-Sterman function~$F_{FT}(x,x)$~\cite{Aybat:2011ge}; 
and $H_{1}^\perp(z,z^2p_\perp^2)$ to its  first $p_{\perp}$-moment~\cite{Kang:2015msa}, defined as
\begin{equation}
H_1^{\perp(1)}(z)
    \equiv z^2\!\int \!d^2 \vec{p}_\perp \frac{p_\perp^2} {2M_h^2}
    H_{1}^\perp(z,z^2p_\perp^2) \, ,
\end{equation}
where $M_h$ is the hadron mass and $p_\perp$ the parton transverse momentum. Note $H_1^{\perp(1)}(z)$ is a CT3 function (the so-called kinematical type~\cite{Kanazawa:2015ajw}).

The same set of functions, $h_1(x)$, $F_{FT}(x,x)$,
$H_1^{\perp(1)}(z)$ in the OPE of TMDs are also the
non-perturbative objects that drive the collinear SSA 
$A_N^h$ in $p^\uparrow p\to h\,X$~\cite{Qiu:1998ia, Kouvaris:2006zy, Koike:2009ge, Kang:2011hk, Metz:2012ct, Beppu:2013uda}.
In fact, in the CT3 framework, the main cause of $A_N^h$ can be
explained by the coupling of $h_1(x)$ to $H_1^{\perp(1)}(z)$ and another
multi-parton correlator $\tilde{H}(z)$~\cite{Kanazawa:2014dca,Gamberg:2017gle}.   
The latter generates the $P_{hT}$-integrated SIDIS
$A_{UT}^{\sin\phi_S}$ asymmetry by coupling with $h_1(x)$~\cite{Bacchetta:2006tn}.  In $A_N^h$ we include both the Qiu-Sterman (``Sivers-type'') and fragmentation (``Collins-type'') terms in our analysis.  As in Refs.~\cite{Kanazawa:2014dca,Gamberg:2017gle}, we again find the former is negligible while the latter is dominant.  Based on the above discussion, one can argue that SSAs  have a common origin, namely,
multi-parton correlations.  

We present, for the first time, a phenomenological verification
of this by performing a simultaneous QCD global analysis of 
$A_{\rm SIDIS}^{\rm Siv}$, 
$A_{\rm SIDIS}^{\rm Col}$, 
$A_{\rm DY}^{\rm Siv}$, 
$A_{\rm SIA}^{\rm Col}$, 
and $A_N^h$. In addition, the fact that we are able to describe both $A_{\rm SIDIS}^{\rm Siv}$ and $A_N^h$ (where the latter includes both ``Collins-type'' and ``Sivers-type'' contributions) further indicates a resolution to the ``sign-mismatch'' puzzle between the Sivers function and Qiu-Sterman function~\cite{Kang:2011hk} found when using the parton model 
relation Eq.~\eqref{eq:relation}. 

\begin{table*}
\centering
\begin{tabular}{ |c|c|c|c|c| }
 \hline
{\bf Observable} & 
{\bf Reactions} & 
{\bf Non-Perturbative Function(s)} &  
$\boldsymbol{\chi^2}/\boldsymbol{N_{\rm pts.}}$ &
{\bf Refs.} 
\\\hline \hline

$A_{\rm SIDIS}^{\rm Siv}$  &
$e+(p,d)^\uparrow\to e+(\pi^+,\pi^-,\pi^0)+X$  & 
$f^{\perp}_{1T}(x,k_T^2)$  & 
$150.0/126=1.19$ & 
\cite{Airapetian:2009ae, Alekseev:2008aa,Adolph:2014zba}  
\\

$A_{\rm SIDIS}^{\rm Col}$  &
$e+(p,d)^\uparrow\to e+(\pi^+,\pi^-,\pi^0)+X$  & 
$h_{1}(x,k_T^2),H_1^{\perp}(z,z^2 p_\perp^2)$  & 
$111.3/126=0.88$ & 
\cite{Airapetian:2010ds, Alekseev:2008aa,Adolph:2014zba} 
\\\hline

$A_{\rm SIA}^{\rm Col}$  &
$e^++e^-\to \pi^+\pi^-(UC,UL)+X$  & 
$H_1^{\perp}(z,z^2 p_\perp^2)$  & 
$154.5/176=0.88$ & 
\cite{Seidl:2008xc,TheBABAR:2013yha,Aubert:2015hha,Ablikim:2015pta} 
\\\hline

$A_{\rm DY}^{\rm Siv}$  &
$\pi^-\!+p^\uparrow\to \mu^+\mu^-+X$  & 
$f_{1T}^{\perp}(x,k_T^2)$  & 
$5.96/12=0.50$ & 
\cite{Aghasyan:2017jop} 
\\

$A_{\rm DY}^{\rm Siv}$  &
$p^\uparrow +p \to (W^+,W^-,Z)+X$  & 
$f_{1T}^{\perp}(x,k_T^2)$  & 
$31.8/17=1.87$ & 
\cite{Adamczyk:2015gyk} 
\\
 
$A_N^h$  &
$p^\uparrow + p\to (\pi^+,\pi^-,\pi^0) + X$  & 
$h_1(x),F_{FT}(x,x)= \tfrac{1} {\pi}f_{1T}^{\perp(1)}(x),H_1^{\perp(1)}(z)$  & 
$66.5/60=1.11$ & 
\cite{Lee:2007zzh,Adams:2003fx,Abelev:2008af,Adamczyk:2012xd} 
\\[0.05cm]
\hline 
\end{tabular}
\caption{
    Summary of the SSAs analyzed in our global fit.  There are a
    total of 18 different reactions. (UC and UL stand for ``unlike-charged'' and ``unlike-like'' pion combinations.) 
    There are also a total of 6 non-perturbative functions when one takes into account flavor separation.   \vspace{-0.3cm}} 
\label{t:sum}
\end{table*}

We further claim that such an analysis 
serves as a universality test since
1)~The system must be over-constrained, i.e., the
number of equations relating partonic functions to observables must be
larger than the number of partonic functions.~2)~Each function
must appear at least twice in such equations.~3)~There must be
reasonable kinematical overlap between observables. 
These conditions are satisfied in our analysis,
as summarized in Table~\ref{t:sum}.  
There is also considerable kinematical overlap in $x$, $z$, and $Q^2$
between observables.  
SIDIS  covers a region 
$x\lesssim 0.3$, $0.2\lesssim z\lesssim 0.6$, 
and $2 \lesssim Q^2\lesssim 40\,{\rm GeV^2}$.  
SIA data has $0.2\lesssim z\lesssim 0.8$ 
and $Q^2\approx 13\,{\rm GeV^2}$ 
or $110\, {\rm GeV^2}$. 
For DY data, $0.1\lesssim x\lesssim 0.35$ 
and $Q^2\approx 30\,{\rm GeV^2}$ or $(80\,{\rm GeV})^2$.  
Lastly, $A_N^h$ integrates from $x_{min}$ to 1 and $z_{min}$ to 1.  For $A_N^{\pi^\pm}$ data from BRAHMS, $0.2\lesssim (x_{min}, z_{min})\lesssim 0.3$, 
with  $1\lesssim Q^2\lesssim 6\,{\rm GeV^2}$. The  $A_N^{\pi^0}$ data from STAR has $0.2\lesssim (x_{min}, z_{min})\lesssim 0.7$, 
and  $1\lesssim Q^2\lesssim 13\,{\rm GeV^2}$.  
Moreover, we provide additional evidence in Sec.~\ref{s:pheno} that SSAs for TMD and CT3 observables have a common origin by first extracting the TMDs from {\it only SSAs in SIDIS, DY, and $e^+e^-$}
and then making {\it predictions} for $A_N^\pi$ based on those results.  A necessary condition for TMD and CT3 SSAs to have the same dynamical origin is that, within error bands, our predictions should describe the $A_N^\pi$ measurements.  Indeed, this is exactly what we find, as we will show later in Sec.~\ref{s:pheno}.

\section{Methodology \label{s:method}}
To perform our global analysis, we must postulate a
functional form for the non-perturbative functions. 
For the TMDs, we
 decouple the $x$ and $k_T$ ($z$ and $p_\perp$) dependence.  This is phenomenologically well motivated  within the literature and has been successfully used in a wide variety of reactions -- see, e.g.,
Refs.~\cite{Anselmino:2005nn,Anselmino:2000vs,Anselmino:2005ea,Vogelsang:2005cs,Collins:2005ie,Collins:2005rq,Anselmino:2007fs,Anselmino:2008jk,Schweitzer:2010tt,Qiu:2011ai,Anselmino:2013vqa,Signori:2013mda,Anselmino:2013lza,Boer:2014tka,DAlesio:2020wjq,Callos:2020qtu}.  This ansatz is also supported by a lattice QCD calculation in Ref.~\cite{Orginos:2017kos}.  We employ a Gaussian parametrization for the transverse
momentum dependence. This assumes most of the transverse momentum is non-perturbative and thus related to intrinsic properties of the colliding hadrons rather than to hard gluon radiation.

Although this type of parametrization does not have the complete features of
TMD evolution,
it was shown in 
Refs.~\cite{Anselmino:2016uie,Anselmino:2015sxa}
that utilizing such a parametrization is comparable to full TMD evolution at next-to-leading-logarithmic accuracy~\cite{Sun:2013dya,Kang:2014zza,Kang:2015msa,Echevarria:2014xaa,Kang:2017btw}.  
In addition, asymmetries are ratios of cross sections 
where evolution and next-to-leading order effects tend to cancel out~\cite{Kang:2017btw}. 
We also implement a DGLAP-type evolution for the collinear twist-3  functions
analogous to Ref.~\cite{Duke:1983gd}, where a
double-logarithmic $Q^2$-dependent term is explicitly added to the
parameters.  For the collinear twist-2 PDFs and FFs, we use the standard leading order DGLAP evolution.

For the unpolarized and transversity TMDs we have
\begin{eqnarray}
f^q(x,k_T^2) 
&=& f^q(x)\ {\cal G}_{f}^q(k_T^2)\,,
\label{eq:f1h1-gauss}
\end{eqnarray}
where the generic function $f = f_1$ or $h_1$, and
\begin{eqnarray}
{\cal G}_f^q(k_T^2)
&=& \frac{1}{\pi\avkT_f^q}\;
{\exp\left[{-\frac{\kT^2}{\avkT_f^q}}\right]}.
\label{eq:Ggauss}
\end{eqnarray}
Using the relation  
$\pi F_{FT}(x,x)=f_{1T}^{\perp(1)}(x)$~\cite{Boer:2003cm}, 
the Sivers function reads
\begin{eqnarray}
f_{1T}^{\perp \,q}(x,k_T^2)
&=& \frac{2 M^2}{\avkT^{q}_{f_{1T}^\perp}}\,
    \pi F_{FT}(x,x)\
    {\cal G}_{f_{1T}^\perp}^{q}\!(k_T^2)\,.
\label{e:sivers}
\end{eqnarray}

For the TMD FFs, the unpolarized function is parametrized as
\begin{eqnarray}
D_1^{h/q}(z,z^2\pperp^2)
&=& D_1^{h/q}(z)\ {\cal G}_{D_1}^{h/q}(z^2\pperp^2)\,,
\end{eqnarray}
while the Collins FF reads
\begin{eqnarray}
H_1^{\perp h/q}(z,z^2\pperp^2)
&=& \frac{2 z^2 M_h^2}{\avpperp^{h/q}_{H_1^\perp}}\,
    H_{1\, h/q}^{\perp (1)}(z)\
    {\cal G}_{H_1^\perp}^{h/q}(z^2\pperp^2)\,,
\label{e:collins}
\end{eqnarray}
where we have explicitly written its $z$ dependence in terms of its
first moment $H_{1\, h/q}^{\perp (1)}(z)$~\cite{Kang:2015msa}. The widths for the FFs are denoted as $\avpperp^{h/q}_{D}$, where $D=D_1\,{\rm or}\, H_1^\perp$. (Note that the hadron transverse momentum $\vec{P}_\perp = -z\vec{p}_\perp$.) 
For $f_1^q(x)$ and $D_1^q(z)$ we use the leading order
CJ~\cite{Accardi:2016qay} and DSS~\cite{deFlorian:2007ekg}
functions.  
The pion PDFs  are taken from Ref.~\cite{Barry:2018ort} and are next-to-leading order~\footnote{The precision of the COMPASS Drell-Yan data is such that using next-to-leading order pion PDFs will not affect our results.}.

Note Eqs.~(\ref{eq:f1h1-gauss}), (\ref{e:sivers}), (\ref{e:collins})
make clear that the underlying non-perturbative functions, 
    $h_1(x)$, 
    $F_{FT}(x,x)$, 
    $H_1^{\perp(1)}(z)$, 
that drive the (TMD) SSAs 
    $A_{\rm SIDIS}^{\rm Siv}$, 
    $A_{\rm SIDIS}^{\rm Col}$, 
    $A_{\rm DY}^{\rm Siv}$, 
    and $A_{\rm SIA}^{\rm Col}$, 
are the same collinear functions that enter the SSA $A_N^h$ (along with
$\tilde{H}(z)$).  
We generically parametrize these collinear functions as
\begin{equation}
F^q(x)\!=\!\frac{N_q\,x^{a_q}(1-x)^{b_q}(1+\gamma_q\,x^{\alpha_q}(1-x)^{\beta_q})}
            {{\rm B}[a_q\!+\!2,b_q\!+\!1]
             +\gamma_q {\rm B}[a_q\!
             +\!\alpha_q\!+\!2,b_q\!
             +\!\beta_q\!+\!1]}   
\,, 
\label{e:colfunc}
\end{equation}
where $F^q=h_1^q,\pi F_{FT}^q$, $H_{1 \,h/q}^{\perp (1)}$ (with $x\to
z$ for the Collins function), and $B$ is the Euler beta function.  
In the course of our analysis, we found that $\tilde{H}(z)$ was
consistent with zero within error bands.  
Therefore, data on the aforementioned ($P_{hT}$-integrated)
$A_{UT}^{\sin\phi_S}$ asymmetry in SIDIS is needed to properly
constrain $\tilde{H}(z)$. 
For now, we set $\tilde{H}(z)$ to zero, which is consistent with
preliminary data from HERMES~\cite{Schnell:2010zza} and COMPASS~\cite{Parsamyan:2013fia} showing a 
small
$A_{UT}^{\sin\phi_S}$.

For the collinear PDFs $h_1^q(x)$ and $\pi F_{FT}^q(x,x)$, we only
allow $q=u,d$ and set antiquark functions to zero.  
For both functions, $\{\gamma,\alpha,\beta\}$ are not used,  and we  set $b_u=b_d$.  This approach is similar to previous analyses~\cite{Anselmino:2013vqa,
       Echevarria:2014xaa,
       Kang:2015msa,
       Anselmino:2015sxa,
       Anselmino:2016uie,
       DAlesio:2020vtw}.
For the collinear FF $H_{1 \,h/q}^{\perp (1)}(z)$, we allow for
favored ($fav$)  and unfavored ($unf$) parameters. 
We also found that, similar to what
has been done in fits of unpolarized collinear
FFs~\cite{deFlorian:2007ekg}, $\{\gamma, \beta\}$ are
needed for $H_{1 \,h/q}^{\perp (1)}(z)$, while $\alpha$ can be set to zero since $A_{\rm SIA}^{\rm Col}$, $A_{\rm SIDIS}^{\rm Col}$ are at
$z\gtrsim 0.2$. The need for $\{\gamma, \beta\}$ is due to the fact that the
data for $A_{\rm SIA}^{\rm Col}$ has a different shape at smaller
versus larger $z$.  Indeed, we found that $(\chi^2/N_{\rm pts.})_{\rm SIA} = 3.85$ if $H_{1 \,h/q}^{\perp (1)}(z)$ only has a functional form proportional to $Nz^{a}(1-z)^{b}$.
In the end we have a total of 20 parameters for the collinear functions.  
There are also 4 parameters for the transverse momentum widths associated with
$h_1$, $f_{1T}^\perp$, and $H_{1}^{\perp}$:
    $\avkT^{u}_{f_{1T}^\perp}=\avkT^{d}_{f_{1T}^\perp}
                              \equiv\avkT_{f_{1T}^\perp}$; 
    $\avkT_{h_1}^u=\avkT_{h_1}^d\equiv \avkT_{h_1}$; 
    $\avpperp^{fav}_{H_1^\perp}$ and
    $\avpperp^{unf}_{H_1^\perp}$.
    
We extract unpolarized TMD widths~\cite{Anselmino:2005nn,Signori:2013mda,Anselmino:2013lza} by including HERMES
pion and kaon multiplicities~\cite{Airapetian:2012ki}, which involves 6 more parameters:~$\avkT^{val}_{f_1}$,
$\avkT^{sea}_{f_1}$, $\avpperp^{fav}_{D_1^{\{\pi,K\}}}$,
$\avpperp^{unf}_{D_1^{\{\pi,K\}}}$.~The pion PDF widths are taken to be the same as those for the proton. We also include normalization parameters for each data set to account for correlated systematic uncertainties.

We use the multi-step strategy in a Monte Carlo framework developed in Ref.~\cite{Sato:2019yez} to reliably sample the Bayesian posterior distribution for the parameters.
This approach allows us to determine the relevant regions in parameter space, and give state-of-the-art uncertainty quantification, for the hadronic structures that best describe the data.

\section{Phenomenological Results \label{s:pheno}}
\begin{figure}[t]
\includegraphics[width=0.425\textwidth]{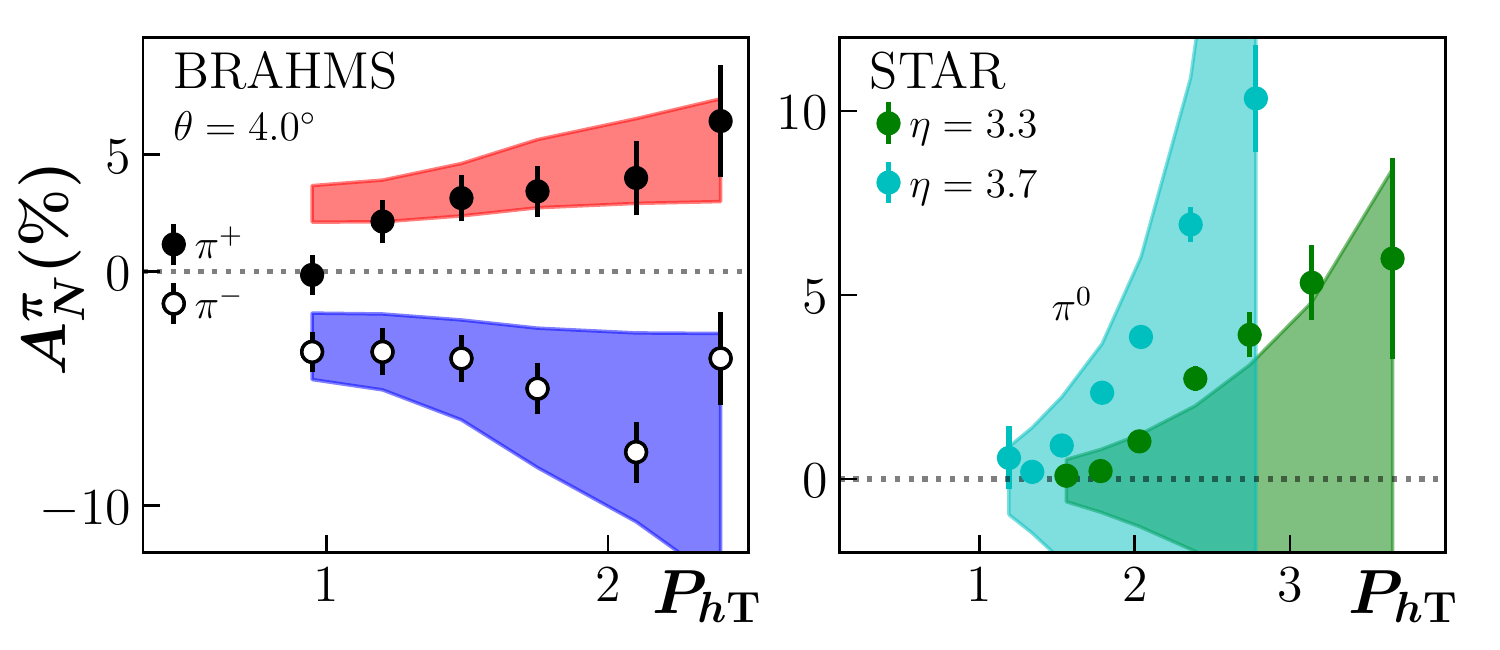}  \vspace{-0.5cm}
\caption{
  Predictions for $A_{N}^{\pi}$ using TMDs extracted from only $A_{\rm SIDIS}^{\rm Siv}$, 
    $A_{\rm SIDIS}^{\rm Col}$, 
    $A_{\rm DY}^{\rm Siv}$, 
    and $A_{\rm SIA}^{\rm Col}$. Similar results are found for the other BRAHMS and STAR data sets. 
}
\label{f:an_pred}
\end{figure}

 We first test the universality of our proposed mechanism by making {\it predictions} for $A_N^\pi$ using TMDs extracted from {\it only SSAs in SIDIS, DY, and $e^+e^-$}. 
 The results are shown in Fig.~\ref{f:an_pred} and are similar to what was found in Ref.~\cite{Gamberg:2017gle}.  
 As one can see, both the BRAHMS and STAR data fall within the theoretical predictions. The large uncertainties of the STAR predictions are due to the fact that the $x$-dependent PDFs (transversity and Qiu-Sterman) must be extrapolated beyond where they are constrained by the TMD SSAs.  
 By including $A_N^\pi$ data in a simultaneous QCD global analysis of SSAs, we can decrease the theoretical error bands and isolate the PDF and FF solutions that optimize the description of all measurements.  

 We also emphasize that the number of parameters and functional form used in this fit, as described in Sec.~\ref{s:method}, do {\it not} guarantee one would be able to successfully describe all SSA data simultaneously.  In general, we are interested in whether certain functions (transversity, Qiu-Sterman, Collins first moment) have universal values for a given kinematic point irrespective of the process in  which they are used.  The answer to this question should be independent of how the functions are parametrized.  In addition, if our parametrization was too flexible to where we overfit the data, one would expect poor predictions for $A_N^\pi$ in Fig.~\ref{f:an_pred}, which is not the case. Note that if the $A_N^\pi$ data did not fall within the predictions of Fig.~\ref{f:an_pred}, one would not expect to simultaneously describe all SSA data.  We stress no additional parameters are introduced when $A_N^\pi$ is included in the combined analysis with TMD SSAs.
\begin{figure}[h]
\includegraphics[width=0.495\textwidth]{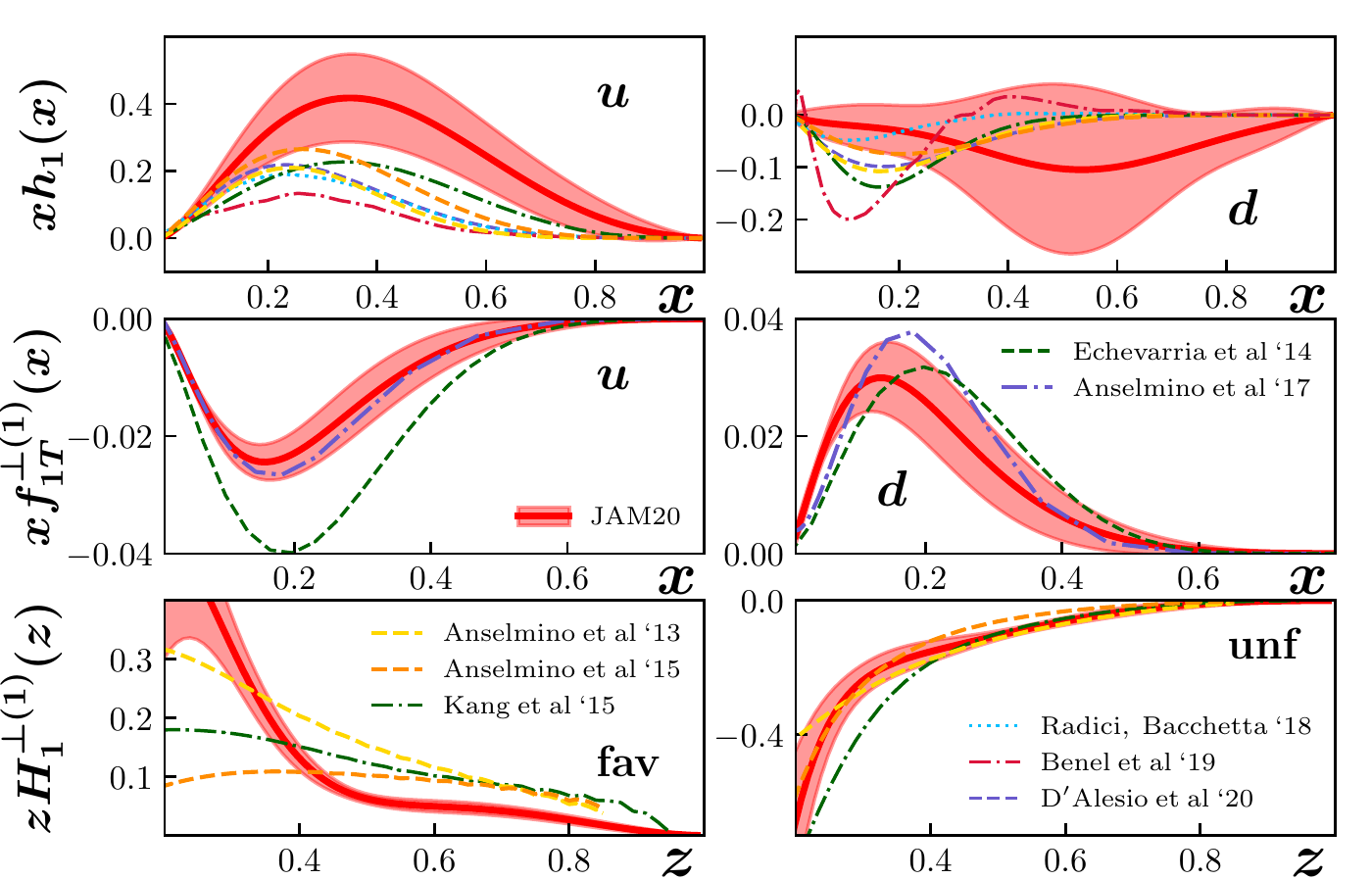}\vspace{-0.2cm}
\caption{The extracted functions $h_1(x)$, $f_{1T}^{\perp(1)}(x)$, and
$H_1^{\perp(1)}(z)$ at $Q^2=4$ GeV$^2$ from our (JAM20) global
analysis (red solid curves with 1-$\sigma$ CL error bands).  The
functions from other groups
~\cite{Anselmino:2013vqa,
       Echevarria:2014xaa,
       Kang:2015msa,
       Anselmino:2015sxa,
       Anselmino:2016uie,
       Radici:2018iag,
       Benel:2019mcq,
       DAlesio:2020vtw} 
are also shown.\vspace{-0.25cm}} 
\label{f:qcf}
\end{figure}
\begin{figure}[h]
\includegraphics[width=0.485\textwidth]{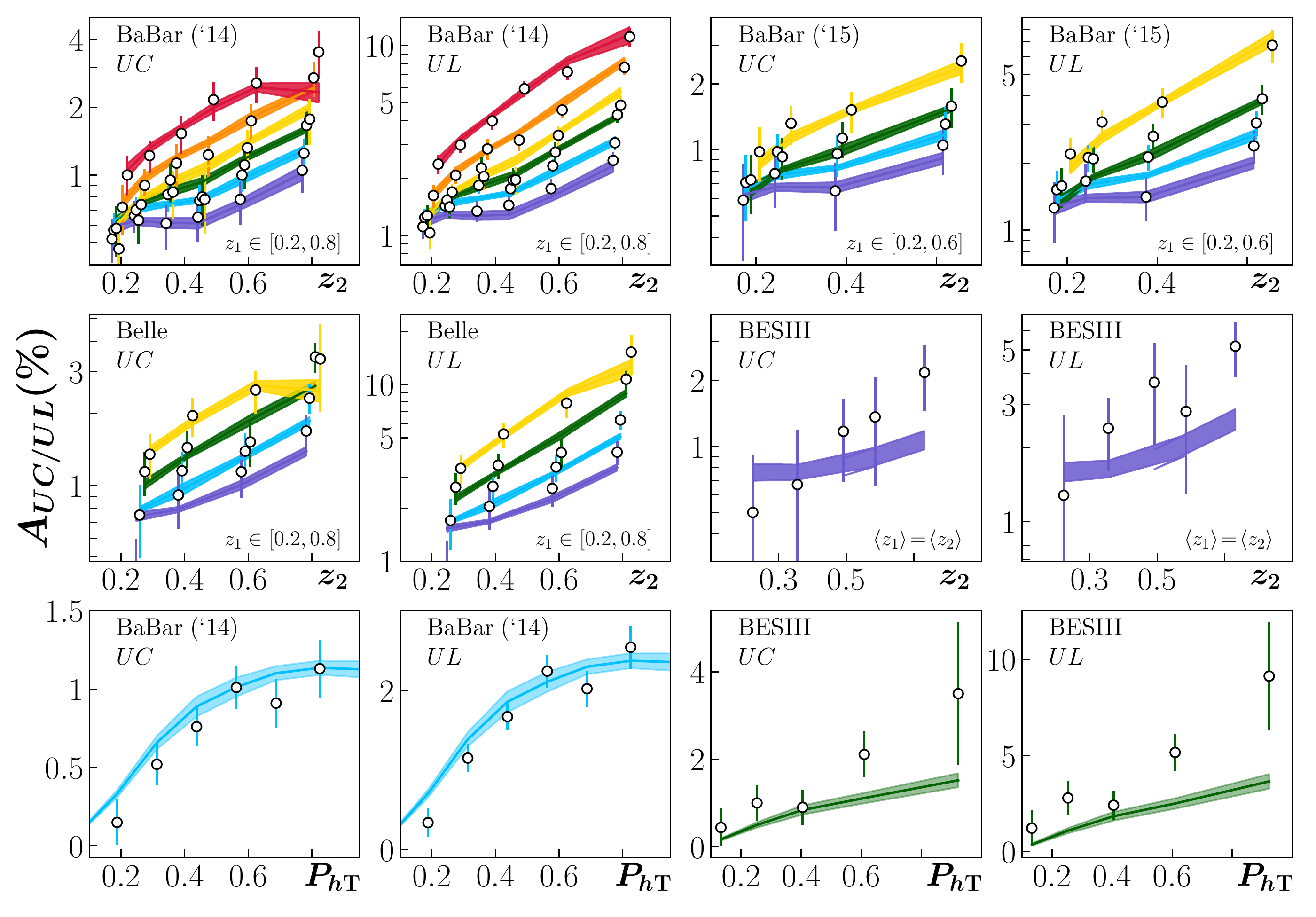}\vspace{-0.4cm}
\caption{
  Theory compared to experiment for $A_{\rm SIA}^{\rm Col}$. 
}
\label{f:thy_vs_data_sia}
\end{figure}
\begin{figure}[h]
\vspace{-0.4cm}
\includegraphics[width=0.485\textwidth]{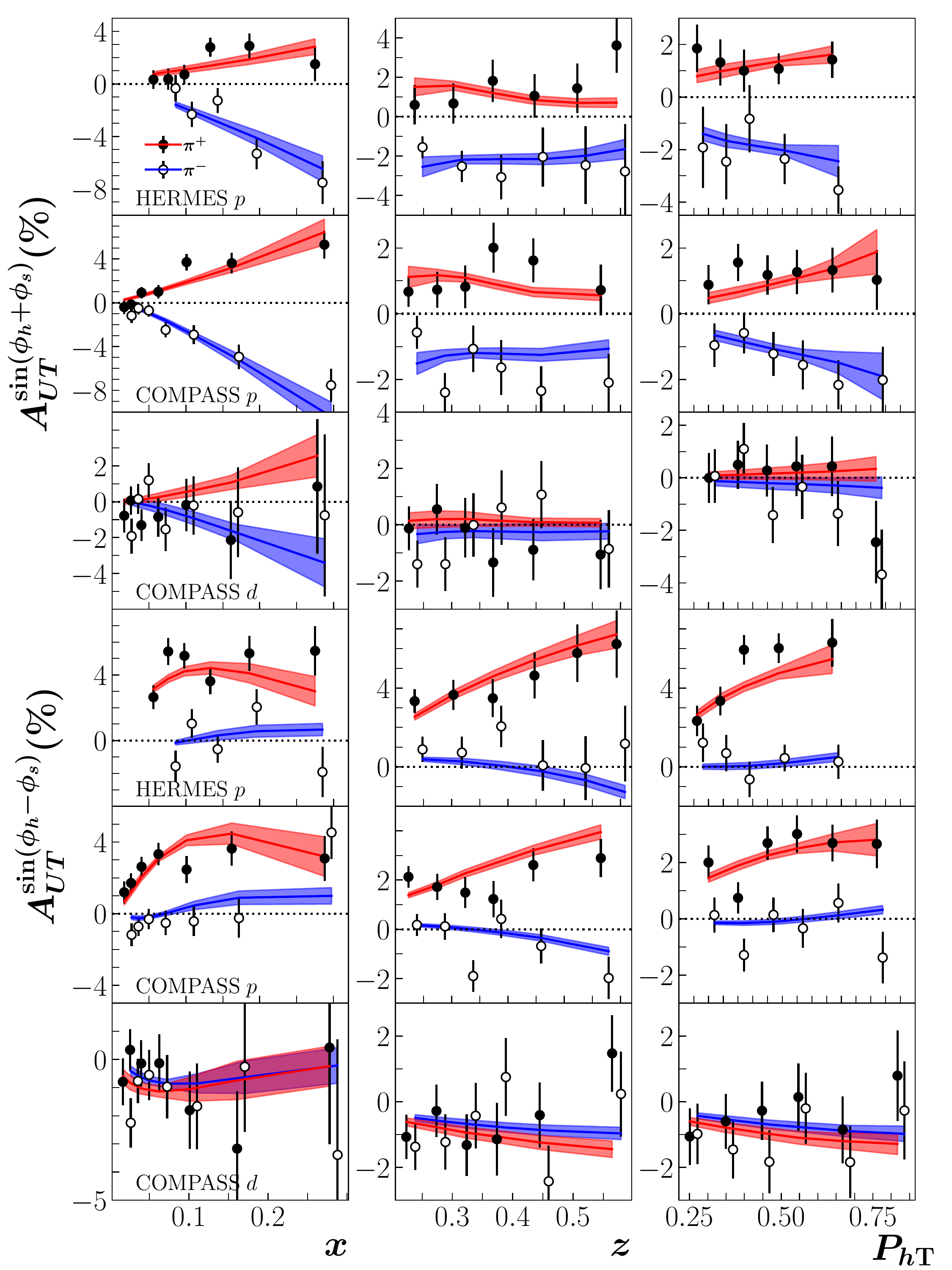}\vspace{-0.35cm}
\caption{Theory compared to experiment for 
$A_{\rm SIDIS}^{\rm Col/Siv}$. 
\vspace{-0.4cm}} 
\label{f:thy_vs_data_sidis}
\end{figure}
\begin{figure}[h]
\centering
\includegraphics[width=0.485\textwidth]{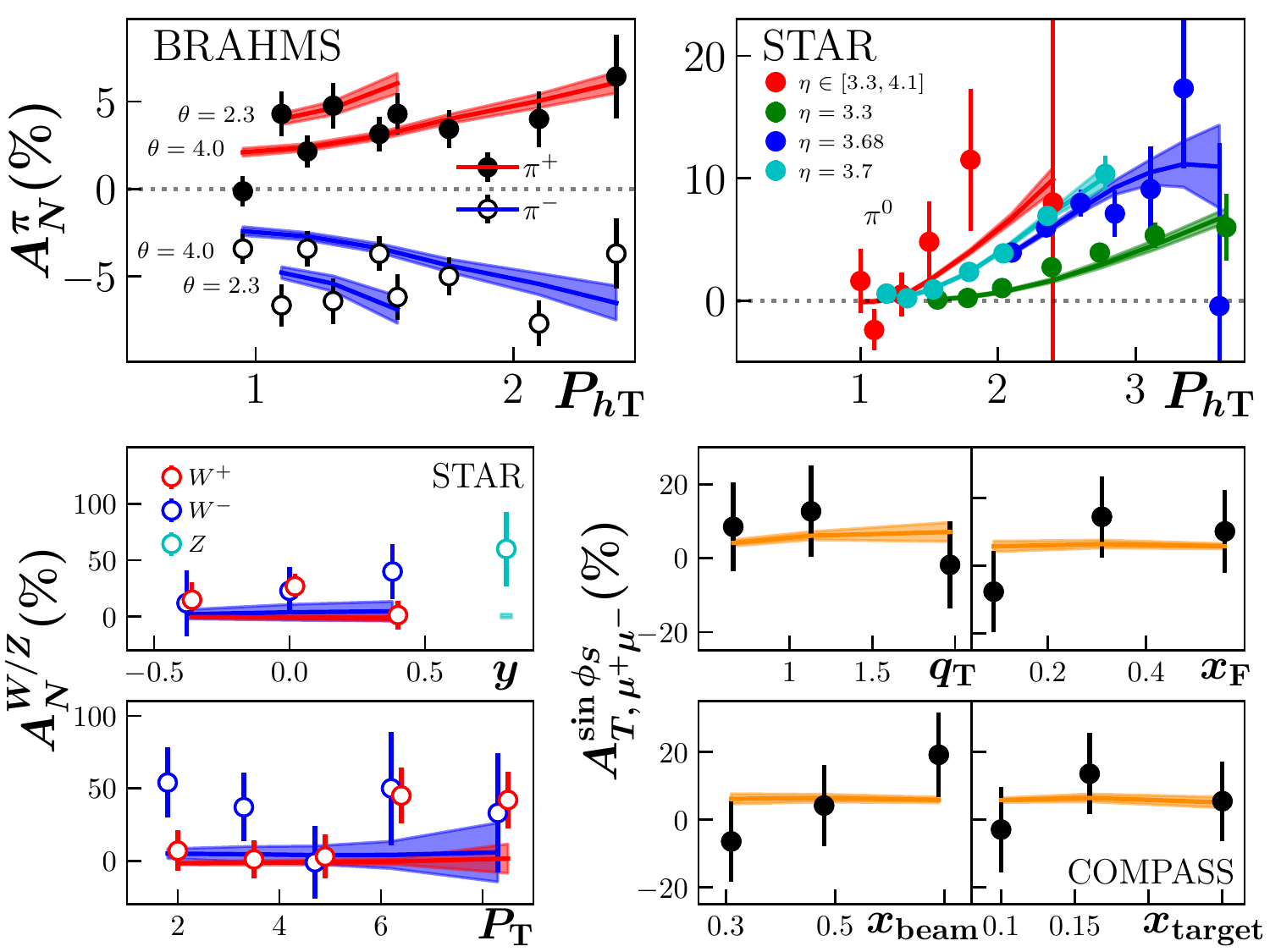}\vspace{-0.3cm}
\caption{Theory compared to experiment for $A_N^\pi\;$
 and $A_{\rm DY}^{\rm Siv}$.
\vspace{-0.3cm}}
\label{f:thy_vs_data_pp}
\end{figure}

We now perform our simultaneous QCD global analysis of the SSA data summarized in Table~\ref{t:sum}.
The standard cuts of 
$0.2 < z< 0.6,\; Q^2>1.63\,{\rm GeV^2}, \;{\rm and}\;0.2 <P_{hT}<0.9 \,{\rm GeV}$ 
have been applied to all SIDIS data sets~\cite{Anselmino:2013lza} and $P_{hT}>1\,{\rm GeV}$ to all $A_N^\pi$ data sets~\cite{Kanazawa:2014dca,Gamberg:2017gle}, giving us a total of 517 SSA data points in the fit along with 807
HERMES multiplicity~\cite{Airapetian:2012ki} data points.  
The extracted functions~\cite{JAM20:code} 
and their comparison to other groups are shown
in Fig.~\ref{f:qcf}. 
We obtain a good agreement between theory and experiment, 
as one sees in Figs.~\ref{f:thy_vs_data_sia}--\ref{f:thy_vs_data_pp}.  
Specifically we find
$(\chi^2/N_{\rm pts.})_{\rm SSA}=520/517=1.01$ for SSA data alone, 
and $\chi^2/N_{\rm pts.}=1373/1324=1.04$ for all data, 
including HERMES multiplicities.
\begin{figure}[t]
\centering
\includegraphics[width=0.495\textwidth]{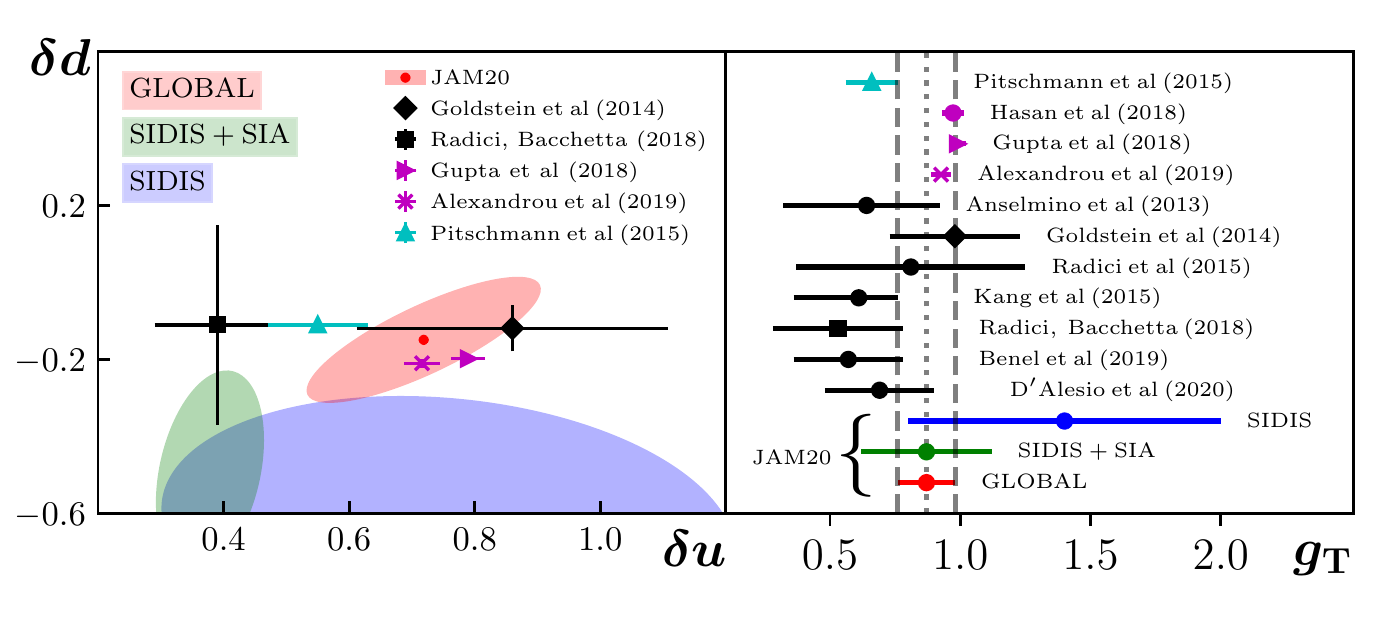}\vspace{-0.5cm}
\caption{
The tensor charges $\delta u$, $\delta d$, and $g_T$. Our (JAM20) results  at
$Q^2=4$~GeV$^2$ along with others from phenomenology (black), lattice QCD (purple), and
Dyson-Schwinger (cyan).
\vspace{-0.6cm}} 
\label{f:gT}
\end{figure}

Figure~\ref{f:gT} displays our extracted tensor charges of the nucleon.
The individual flavor charges 
$\delta q\!\equiv\!\int_0^1\!dx\,[h_1^q(x)-h_1^{\bar q}(x)]$ 
are shown along with the isovector combination 
$g_T\equiv \delta u- \delta d$.  
We compare our results to those from lattice QCD computations at the
physical point~\cite{Gupta:2018qil,Hasan:2019noy,Alexandrou:2019brg},
other phenomenological extractions~\cite{Anselmino:2013vqa,
       Goldstein:2014aja,
       Radici:2015mwa,
       Kang:2015msa,
       Radici:2018iag,
       Benel:2019mcq,
       DAlesio:2020vtw},
and a calculation using Dyson-Schwinger
equations~\cite{Pitschmann:2014jxa}.  
From Fig.~\ref{f:gT}, the strong impact
of including more SSA data sets is clear, highlighting the importance of carrying out 
a simultaneous extraction of partonic functions in a global analysis.
In going from 
${\rm SIDIS}\to{\rm (SIDIS+SIA)}\to{\rm GLOBAL}$ (where GLOBAL in particular includes $A_N^\pi$),
we find $g_T=1.4(6)\to 0.87(25)\to 0.87(11)$.  
This is the most precise phenomenological determination of $g_T$ to date.
All of the inferred tensor charges 
($\delta u$, $\delta d$, and $g_T$) are in excellent agreement with
lattice QCD data.  
As can be seen from Fig.~~\ref{f:gT}, including $A_N^\pi$ is crucial to achieve the agreement between our results 
$\delta u =0.72(19),\, \delta d=-0.15(16)$ and those from lattice QCD.  

\section{Conclusions} 
In this paper we have performed the first simultaneous QCD global analysis of the
available SSA data in SIDIS, DY, $e^+e^-$ annihilation, and
proton-proton collisions. 
The predictive power exhibited by the results of the combined analysis indicates SSAs have a common origin.
Namely, they are due to the intrinsic quantum-mechanical interference
from multi-parton states. 
Our findings imply that the
effects are predominantly non-perturbative and intrinsic to hadronic
wavefunctions. 
Also, the extracted up and down quark tensor charges
are in excellent agreement with lattice QCD.

The future data from JLab-12~GeV~\cite{Dudek:2012vr}, COMPASS~\cite{Bradamante:2018ick, Gautheron:2010wva},
an upgraded RHIC~\cite{Aschenauer:2015eha}, Belle II~\cite{Kou:2018nap}, and the Electron-Ion Collider~\cite{Boer:2011fh,Accardi:2012qut} will help to
reduce the uncertainties of the extracted functions. Measurements that have kinematical overlap to the current data, like SIDIS data from JLab-12~GeV~\cite{Dudek:2012vr,Chen:2014psa} and an EIC~\cite{Accardi:2012qut}, more precise Drell-Yan data from COMPASS~\cite{Gautheron:2010wva,Bradamante:2018ick} and STAR~\cite{Aschenauer:2015eha}, and new $A_N$ and pion-in-jet data from STAR~\cite{Aschenauer:2015eha}, will test our results.   Ultimately, all these measurements will lead to a deeper understanding of hadronic structure.

\section{Acknowledgments}  
This work has been supported by the NSF under Grants 
  No.~PHY-2012002 (A.P.), 
  No.~PHY-1720486 (Z.K.) and No.~PHY-1945471 (Z.K.), No.~PHY-2011763 (D.P.), the U.S. Department of Energy, under
contracts 
  No. DE-FG02-07ER41460 (L.G.), 
  No.~DE-AC05-06OR23177 (A.P., N.S., T.R.) under  which  Jefferson  Science  Associates,  LLC,  manages and operates Jefferson Lab, 
a Lebanon Valley College (LVC) Arnold Student-Faculty Research Grant (J.A.M.~and D.P.), 
and within the framework of the TMD Topical Collaboration.


\begin{thebibliography}{127}%
\makeatletter
\providecommand \@ifxundefined [1]{%
 \@ifx{#1\undefined}
}%
\providecommand \@ifnum [1]{%
 \ifnum #1\expandafter \@firstoftwo
 \else \expandafter \@secondoftwo
 \fi
}%
\providecommand \@ifx [1]{%
 \ifx #1\expandafter \@firstoftwo
 \else \expandafter \@secondoftwo
 \fi
}%
\providecommand \natexlab [1]{#1}%
\providecommand \enquote  [1]{``#1''}%
\providecommand \bibnamefont  [1]{#1}%
\providecommand \bibfnamefont [1]{#1}%
\providecommand \citenamefont [1]{#1}%
\providecommand \href@noop [0]{\@secondoftwo}%
\providecommand \href [0]{\begingroup \@sanitize@url \@href}%
\providecommand \@href[1]{\@@startlink{#1}\@@href}%
\providecommand \@@href[1]{\endgroup#1\@@endlink}%
\providecommand \@sanitize@url [0]{\catcode `\\12\catcode `\$12\catcode
  `\&12\catcode `\#12\catcode `\^12\catcode `\_12\catcode `\%12\relax}%
\providecommand \@@startlink[1]{}%
\providecommand \@@endlink[0]{}%
\providecommand \url  [0]{\begingroup\@sanitize@url \@url }%
\providecommand \@url [1]{\endgroup\@href {#1}{\urlprefix }}%
\providecommand \urlprefix  [0]{URL }%
\providecommand \Eprint [0]{\href }%
\providecommand \doibase [0]{http://dx.doi.org/}%
\providecommand \selectlanguage [0]{\@gobble}%
\providecommand \bibinfo  [0]{\@secondoftwo}%
\providecommand \bibfield  [0]{\@secondoftwo}%
\providecommand \translation [1]{[#1]}%
\providecommand \BibitemOpen [0]{}%
\providecommand \bibitemStop [0]{}%
\providecommand \bibitemNoStop [0]{.\EOS\space}%
\providecommand \EOS [0]{\spacefactor3000\relax}%
\providecommand \BibitemShut  [1]{\csname bibitem#1\endcsname}%
\let\auto@bib@innerbib\@empty
\bibitem [{\citenamefont {Kane}\ \emph {et~al.}(1978)\citenamefont {Kane},
  \citenamefont {Pumplin},\ and\ \citenamefont {Repko}}]{Kane:1978nd}%
  \BibitemOpen
  \bibfield  {author} {\bibinfo {author} {\bibfnamefont {G.~L.}\ \bibnamefont
  {Kane}}, \bibinfo {author} {\bibfnamefont {J.}~\bibnamefont {Pumplin}}, \
  and\ \bibinfo {author} {\bibfnamefont {W.}~\bibnamefont {Repko}},\ }\href
  {\doibase 10.1103/PhysRevLett.41.1689} {\bibfield  {journal} {\bibinfo
  {journal} {Phys. Rev. Lett.}\ }\textbf {\bibinfo {volume} {41}},\ \bibinfo
  {pages} {1689} (\bibinfo {year} {1978})}\BibitemShut {NoStop}%
\bibitem [{\citenamefont {Bunce}\ \emph {et~al.}(1976)\citenamefont {Bunce}
  \emph {et~al.}}]{Bunce:1976yb}%
  \BibitemOpen
  \bibfield  {author} {\bibinfo {author} {\bibfnamefont {G.}~\bibnamefont
  {Bunce}} \emph {et~al.},\ }\href@noop {} {\bibfield  {journal} {\bibinfo
  {journal} {Phys. Rev. Lett.}\ }\textbf {\bibinfo {volume} {36}},\ \bibinfo
  {pages} {1113} (\bibinfo {year} {1976})}\BibitemShut {NoStop}%
\bibitem [{\citenamefont {Klem}\ \emph {et~al.}(1976)\citenamefont {Klem} \emph
  {et~al.}}]{Klem:1976ui}%
  \BibitemOpen
  \bibfield  {author} {\bibinfo {author} {\bibfnamefont {R.~D.}\ \bibnamefont
  {Klem}} \emph {et~al.},\ }\href {\doibase 10.1103/PhysRevLett.36.929}
  {\bibfield  {journal} {\bibinfo  {journal} {Phys. Rev. Lett.}\ }\textbf
  {\bibinfo {volume} {36}},\ \bibinfo {pages} {929} (\bibinfo {year}
  {1976})}\BibitemShut {NoStop}%
\bibitem [{\citenamefont {Adams}\ \emph {et~al.}(1991)\citenamefont {Adams}
  \emph {et~al.}}]{Adams:1991rw}%
  \BibitemOpen
  \bibfield  {author} {\bibinfo {author} {\bibfnamefont {D.~L.}\ \bibnamefont
  {Adams}} \emph {et~al.} (\bibinfo {collaboration} {E581}),\ }\href {\doibase
  10.1016/0370-2693(91)91351-U} {\bibfield  {journal} {\bibinfo  {journal}
  {Phys. Lett.}\ }\textbf {\bibinfo {volume} {B261}},\ \bibinfo {pages} {201}
  (\bibinfo {year} {1991})}\BibitemShut {NoStop}%
\bibitem [{\citenamefont {Krueger}\ \emph {et~al.}(1999)\citenamefont {Krueger}
  \emph {et~al.}}]{Krueger:1998hz}%
  \BibitemOpen
  \bibfield  {author} {\bibinfo {author} {\bibfnamefont {K.}~\bibnamefont
  {Krueger}} \emph {et~al.},\ }\href {\doibase 10.1016/S0370-2693(99)00677-2}
  {\bibfield  {journal} {\bibinfo  {journal} {Phys. Lett.}\ }\textbf {\bibinfo
  {volume} {B459}},\ \bibinfo {pages} {412} (\bibinfo {year}
  {1999})}\BibitemShut {NoStop}%
\bibitem [{\citenamefont {Allgower}\ \emph {et~al.}(2002)\citenamefont
  {Allgower} \emph {et~al.}}]{Allgower:2002qi}%
  \BibitemOpen
  \bibfield  {author} {\bibinfo {author} {\bibfnamefont {C.~E.}\ \bibnamefont
  {Allgower}} \emph {et~al.},\ }\href {\doibase 10.1103/PhysRevD.65.092008}
  {\bibfield  {journal} {\bibinfo  {journal} {Phys. Rev.}\ }\textbf {\bibinfo
  {volume} {D65}},\ \bibinfo {pages} {092008} (\bibinfo {year}
  {2002})}\BibitemShut {NoStop}%
\bibitem [{\citenamefont {Adams}\ \emph {et~al.}(2004)\citenamefont {Adams}
  \emph {et~al.}}]{Adams:2003fx}%
  \BibitemOpen
  \bibfield  {author} {\bibinfo {author} {\bibfnamefont {J.}~\bibnamefont
  {Adams}} \emph {et~al.} (\bibinfo {collaboration} {STAR}),\ }\href@noop {}
  {\bibfield  {journal} {\bibinfo  {journal} {Phys.~Rev.~Lett.}\ }\textbf
  {\bibinfo {volume} {92}},\ \bibinfo {pages} {171801} (\bibinfo {year}
  {2004})},\ \Eprint {http://arxiv.org/abs/hep-ex/0310058} {hep-ex/0310058}
  \BibitemShut {NoStop}%
\bibitem [{\citenamefont {Adler}\ \emph {et~al.}(2005)\citenamefont {Adler}
  \emph {et~al.}}]{Adler:2005in}%
  \BibitemOpen
  \bibfield  {author} {\bibinfo {author} {\bibfnamefont {S.~S.}\ \bibnamefont
  {Adler}} \emph {et~al.} (\bibinfo {collaboration} {PHENIX}),\ }\href
  {\doibase 10.1103/PhysRevLett.95.202001} {\bibfield  {journal} {\bibinfo
  {journal} {Phys.~Rev.~Lett.}\ }\textbf {\bibinfo {volume} {95}},\ \bibinfo
  {pages} {202001} (\bibinfo {year} {2005})},\ \Eprint
  {http://arxiv.org/abs/hep-ex/0507073} {arXiv:hep-ex/0507073} \BibitemShut
  {NoStop}%
\bibitem [{\citenamefont {Lee}\ and\ \citenamefont
  {Videbaek}(2007)}]{Lee:2007zzh}%
  \BibitemOpen
  \bibfield  {author} {\bibinfo {author} {\bibfnamefont {J.~H.}\ \bibnamefont
  {Lee}}\ and\ \bibinfo {author} {\bibfnamefont {F.}~\bibnamefont {Videbaek}}
  (\bibinfo {collaboration} {BRAHMS}),\ }\bibfield  {booktitle} {\emph
  {\bibinfo {booktitle} {{Proceedings, 17th International Spin Physics
  Symposium (SPIN06): Kyoto, Japan, October 2-7, 2006}}},\ }\href {\doibase
  10.1063/1.2750837} {\bibfield  {journal} {\bibinfo  {journal} {AIP Conf.
  Proc.}\ }\textbf {\bibinfo {volume} {915}},\ \bibinfo {pages} {533} (\bibinfo
  {year} {2007})}\BibitemShut {NoStop}%
\bibitem [{\citenamefont {Abelev}\ \emph {et~al.}(2008)\citenamefont {Abelev}
  \emph {et~al.}}]{Abelev:2008af}%
  \BibitemOpen
  \bibfield  {author} {\bibinfo {author} {\bibfnamefont {B.~I.}\ \bibnamefont
  {Abelev}} \emph {et~al.} (\bibinfo {collaboration} {STAR}),\ }\href {\doibase
  10.1103/PhysRevLett.101.222001} {\bibfield  {journal} {\bibinfo  {journal}
  {Phys. Rev. Lett.}\ }\textbf {\bibinfo {volume} {101}},\ \bibinfo {pages}
  {222001} (\bibinfo {year} {2008})},\ \Eprint {http://arxiv.org/abs/0801.2990}
  {arXiv:0801.2990 [hep-ex]} \BibitemShut {NoStop}%
\bibitem [{\citenamefont {Arsene}\ \emph {et~al.}(2008)\citenamefont {Arsene}
  \emph {et~al.}}]{Arsene:2008aa}%
  \BibitemOpen
  \bibfield  {author} {\bibinfo {author} {\bibfnamefont {I.}~\bibnamefont
  {Arsene}} \emph {et~al.} (\bibinfo {collaboration} {BRAHMS}),\ }\href
  {\doibase 10.1103/PhysRevLett.101.042001} {\bibfield  {journal} {\bibinfo
  {journal} {Phys. Rev. Lett.}\ }\textbf {\bibinfo {volume} {101}},\ \bibinfo
  {pages} {042001} (\bibinfo {year} {2008})},\ \Eprint
  {http://arxiv.org/abs/0801.1078} {arXiv:0801.1078 [nucl-ex]} \BibitemShut
  {NoStop}%
\bibitem [{\citenamefont {Adamczyk}\ \emph
  {et~al.}(2012{\natexlab{a}})\citenamefont {Adamczyk} \emph
  {et~al.}}]{Adamczyk:2012qj}%
  \BibitemOpen
  \bibfield  {author} {\bibinfo {author} {\bibfnamefont {L.}~\bibnamefont
  {Adamczyk}} \emph {et~al.} (\bibinfo {collaboration} {STAR}),\ }\href
  {\doibase 10.1103/PhysRevD.86.032006} {\bibfield  {journal} {\bibinfo
  {journal} {Phys. Rev.}\ }\textbf {\bibinfo {volume} {D86}},\ \bibinfo {pages}
  {032006} (\bibinfo {year} {2012}{\natexlab{a}})},\ \Eprint
  {http://arxiv.org/abs/1205.2735} {arXiv:1205.2735 [nucl-ex]} \BibitemShut
  {NoStop}%
\bibitem [{\citenamefont {Adamczyk}\ \emph
  {et~al.}(2012{\natexlab{b}})\citenamefont {Adamczyk} \emph
  {et~al.}}]{Adamczyk:2012xd}%
  \BibitemOpen
  \bibfield  {author} {\bibinfo {author} {\bibfnamefont {L.}~\bibnamefont
  {Adamczyk}} \emph {et~al.} (\bibinfo {collaboration} {STAR}),\ }\href
  {\doibase 10.1103/PhysRevD.86.051101} {\bibfield  {journal} {\bibinfo
  {journal} {Phys.~Rev.}\ }\textbf {\bibinfo {volume} {D86}},\ \bibinfo {pages}
  {051101} (\bibinfo {year} {2012}{\natexlab{b}})},\ \Eprint
  {http://arxiv.org/abs/1205.6826} {arXiv:1205.6826 [nucl-ex]} \BibitemShut
  {NoStop}%
\bibitem [{\citenamefont {Bland}\ \emph {et~al.}(2015)\citenamefont {Bland}
  \emph {et~al.}}]{Bland:2013pkt}%
  \BibitemOpen
  \bibfield  {author} {\bibinfo {author} {\bibfnamefont {L.}~\bibnamefont
  {Bland}} \emph {et~al.} (\bibinfo {collaboration} {AnDY}),\ }\href {\doibase
  10.1016/j.physletb.2015.10.001} {\bibfield  {journal} {\bibinfo  {journal}
  {Phys. Lett. B}\ }\textbf {\bibinfo {volume} {750}},\ \bibinfo {pages} {660}
  (\bibinfo {year} {2015})},\ \Eprint {http://arxiv.org/abs/1304.1454}
  {arXiv:1304.1454 [hep-ex]} \BibitemShut {NoStop}%
\bibitem [{\citenamefont {Adare}\ \emph
  {et~al.}(2014{\natexlab{a}})\citenamefont {Adare} \emph
  {et~al.}}]{Adare:2013ekj}%
  \BibitemOpen
  \bibfield  {author} {\bibinfo {author} {\bibfnamefont {A.}~\bibnamefont
  {Adare}} \emph {et~al.} (\bibinfo {collaboration} {PHENIX}),\ }\href
  {\doibase 10.1103/PhysRevD.90.012006} {\bibfield  {journal} {\bibinfo
  {journal} {Phys. Rev.}\ }\textbf {\bibinfo {volume} {D90}},\ \bibinfo {pages}
  {012006} (\bibinfo {year} {2014}{\natexlab{a}})},\ \Eprint
  {http://arxiv.org/abs/1312.1995} {arXiv:1312.1995 [hep-ex]} \BibitemShut
  {NoStop}%
\bibitem [{\citenamefont {Adare}\ \emph
  {et~al.}(2014{\natexlab{b}})\citenamefont {Adare} \emph
  {et~al.}}]{Adare:2014qzo}%
  \BibitemOpen
  \bibfield  {author} {\bibinfo {author} {\bibfnamefont {A.}~\bibnamefont
  {Adare}} \emph {et~al.} (\bibinfo {collaboration} {PHENIX}),\ }\href
  {\doibase 10.1103/PhysRevD.90.072008} {\bibfield  {journal} {\bibinfo
  {journal} {Phys. Rev.}\ }\textbf {\bibinfo {volume} {D90}},\ \bibinfo {pages}
  {072008} (\bibinfo {year} {2014}{\natexlab{b}})},\ \Eprint
  {http://arxiv.org/abs/1406.3541} {arXiv:1406.3541 [hep-ex]} \BibitemShut
  {NoStop}%
\bibitem [{\citenamefont {Airapetian}\ \emph {et~al.}(2014)\citenamefont
  {Airapetian} \emph {et~al.}}]{Airapetian:2013bim}%
  \BibitemOpen
  \bibfield  {author} {\bibinfo {author} {\bibfnamefont {A.}~\bibnamefont
  {Airapetian}} \emph {et~al.} (\bibinfo {collaboration} {HERMES}),\ }\href
  {\doibase 10.1016/j.physletb.2013.11.021} {\bibfield  {journal} {\bibinfo
  {journal} {Phys.~Lett.}\ }\textbf {\bibinfo {volume} {B728}},\ \bibinfo
  {pages} {183} (\bibinfo {year} {2014})},\ \Eprint
  {http://arxiv.org/abs/1310.5070} {arXiv:1310.5070 [hep-ex]} \BibitemShut
  {NoStop}%
\bibitem [{\citenamefont {Allada}\ \emph {et~al.}(2014)\citenamefont {Allada}
  \emph {et~al.}}]{Allada:2013nsw}%
  \BibitemOpen
  \bibfield  {author} {\bibinfo {author} {\bibfnamefont {K.}~\bibnamefont
  {Allada}} \emph {et~al.} (\bibinfo {collaboration} {Jefferson Lab Hall A}),\
  }\href {\doibase 10.1103/PhysRevC.89.042201} {\bibfield  {journal} {\bibinfo
  {journal} {Phys.~Rev.}\ }\textbf {\bibinfo {volume} {C89}},\ \bibinfo {pages}
  {042201} (\bibinfo {year} {2014})},\ \Eprint {http://arxiv.org/abs/1311.1866}
  {arXiv:1311.1866 [nucl-ex]} \BibitemShut {NoStop}%
\bibitem [{\citenamefont {Qiu}\ and\ \citenamefont
  {Sterman}(1991)}]{Qiu:1991pp}%
  \BibitemOpen
  \bibfield  {author} {\bibinfo {author} {\bibfnamefont {J.-W.}\ \bibnamefont
  {Qiu}}\ and\ \bibinfo {author} {\bibfnamefont {G.}~\bibnamefont {Sterman}},\
  }\href@noop {} {\bibfield  {journal} {\bibinfo  {journal} {Phys.~Rev.~Lett.}\
  }\textbf {\bibinfo {volume} {67}},\ \bibinfo {pages} {2264} (\bibinfo {year}
  {1991})}\BibitemShut {NoStop}%
\bibitem [{\citenamefont {Qiu}\ and\ \citenamefont
  {Sterman}(1992)}]{Qiu:1991wg}%
  \BibitemOpen
  \bibfield  {author} {\bibinfo {author} {\bibfnamefont {J.-W.}\ \bibnamefont
  {Qiu}}\ and\ \bibinfo {author} {\bibfnamefont {G.}~\bibnamefont {Sterman}},\
  }\href@noop {} {\bibfield  {journal} {\bibinfo  {journal} {Nucl.~Phys.}\
  }\textbf {\bibinfo {volume} {B378}},\ \bibinfo {pages} {52} (\bibinfo {year}
  {1992})}\BibitemShut {NoStop}%
\bibitem [{\citenamefont {Collins}\ and\ \citenamefont
  {Soper}(1981)}]{Collins:1981uk}%
  \BibitemOpen
  \bibfield  {author} {\bibinfo {author} {\bibfnamefont {J.~C.}\ \bibnamefont
  {Collins}}\ and\ \bibinfo {author} {\bibfnamefont {D.~E.}\ \bibnamefont
  {Soper}},\ }\href {\doibase 10.1016/0550-3213(81)90339-4} {\bibfield
  {journal} {\bibinfo  {journal} {Nucl. Phys.}\ }\textbf {\bibinfo {volume}
  {B193}},\ \bibinfo {pages} {381} (\bibinfo {year} {1981})}\BibitemShut
  {NoStop}%
\bibitem [{\citenamefont {Collins}\ \emph {et~al.}(1985)\citenamefont
  {Collins}, \citenamefont {Soper},\ and\ \citenamefont
  {Sterman}}]{Collins:1984kg}%
  \BibitemOpen
  \bibfield  {author} {\bibinfo {author} {\bibfnamefont {J.~C.}\ \bibnamefont
  {Collins}}, \bibinfo {author} {\bibfnamefont {D.~E.}\ \bibnamefont {Soper}},
  \ and\ \bibinfo {author} {\bibfnamefont {G.}~\bibnamefont {Sterman}},\ }\href
  {\doibase 10.1016/0550-3213(85)90479-1} {\bibfield  {journal} {\bibinfo
  {journal} {Nucl. Phys.}\ }\textbf {\bibinfo {volume} {B250}},\ \bibinfo
  {pages} {199} (\bibinfo {year} {1985})}\BibitemShut {NoStop}%
\bibitem [{\citenamefont {Meng}\ \emph {et~al.}(1996)\citenamefont {Meng},
  \citenamefont {Olness},\ and\ \citenamefont {Soper}}]{Meng:1995yn}%
  \BibitemOpen
  \bibfield  {author} {\bibinfo {author} {\bibfnamefont {R.}~\bibnamefont
  {Meng}}, \bibinfo {author} {\bibfnamefont {F.~I.}\ \bibnamefont {Olness}}, \
  and\ \bibinfo {author} {\bibfnamefont {D.~E.}\ \bibnamefont {Soper}},\ }\href
  {\doibase 10.1103/PhysRevD.54.1919} {\bibfield  {journal} {\bibinfo
  {journal} {Phys. Rev.}\ }\textbf {\bibinfo {volume} {D54}},\ \bibinfo {pages}
  {1919} (\bibinfo {year} {1996})},\ \Eprint
  {http://arxiv.org/abs/hep-ph/9511311} {arXiv:hep-ph/9511311} \BibitemShut
  {NoStop}%
\bibitem [{\citenamefont {Efremov}\ and\ \citenamefont
  {Teryaev}(1982)}]{Efremov:1981sh}%
  \BibitemOpen
  \bibfield  {author} {\bibinfo {author} {\bibfnamefont {A.~V.}\ \bibnamefont
  {Efremov}}\ and\ \bibinfo {author} {\bibfnamefont {O.~V.}\ \bibnamefont
  {Teryaev}},\ }\href@noop {} {\bibfield  {journal} {\bibinfo  {journal}
  {Sov.~J.~Nucl.~Phys.}\ }\textbf {\bibinfo {volume} {36}},\ \bibinfo {pages}
  {140} (\bibinfo {year} {1982})}\BibitemShut {NoStop}%
\bibitem [{\citenamefont {Efremov}\ and\ \citenamefont
  {Teryaev}(1985)}]{Efremov:1984ip}%
  \BibitemOpen
  \bibfield  {author} {\bibinfo {author} {\bibfnamefont {A.}~\bibnamefont
  {Efremov}}\ and\ \bibinfo {author} {\bibfnamefont {O.}~\bibnamefont
  {Teryaev}},\ }\href {\doibase 10.1016/0370-2693(85)90999-2} {\bibfield
  {journal} {\bibinfo  {journal} {Phys.Lett.}\ }\textbf {\bibinfo {volume}
  {B150}},\ \bibinfo {pages} {383} (\bibinfo {year} {1985})}\BibitemShut
  {NoStop}%
\bibitem [{\citenamefont {Qiu}\ and\ \citenamefont
  {Sterman}(1998)}]{Qiu:1998ia}%
  \BibitemOpen
  \bibfield  {author} {\bibinfo {author} {\bibfnamefont {J.-W.}\ \bibnamefont
  {Qiu}}\ and\ \bibinfo {author} {\bibfnamefont {G.}~\bibnamefont {Sterman}},\
  }\href@noop {} {\bibfield  {journal} {\bibinfo  {journal} {Phys.~Rev.}\
  }\textbf {\bibinfo {volume} {D59}},\ \bibinfo {pages} {014004} (\bibinfo
  {year} {1998})},\ \Eprint {http://arxiv.org/abs/hep-ph/9806356}
  {hep-ph/9806356} \BibitemShut {NoStop}%
\bibitem [{\citenamefont {Eguchi}\ \emph {et~al.}(2006)\citenamefont {Eguchi},
  \citenamefont {Koike},\ and\ \citenamefont {Tanaka}}]{Eguchi:2006qz}%
  \BibitemOpen
  \bibfield  {author} {\bibinfo {author} {\bibfnamefont {H.}~\bibnamefont
  {Eguchi}}, \bibinfo {author} {\bibfnamefont {Y.}~\bibnamefont {Koike}}, \
  and\ \bibinfo {author} {\bibfnamefont {K.}~\bibnamefont {Tanaka}},\ }\href
  {\doibase 10.1016/j.nuclphysb.2006.05.036} {\bibfield  {journal} {\bibinfo
  {journal} {Nucl. Phys.}\ }\textbf {\bibinfo {volume} {B752}},\ \bibinfo
  {pages} {1} (\bibinfo {year} {2006})},\ \Eprint
  {http://arxiv.org/abs/hep-ph/0604003} {arXiv:hep-ph/0604003 [hep-ph]}
  \BibitemShut {NoStop}%
\bibitem [{\citenamefont {Kouvaris}\ \emph {et~al.}(2006)\citenamefont
  {Kouvaris}, \citenamefont {Qiu}, \citenamefont {Vogelsang},\ and\
  \citenamefont {Yuan}}]{Kouvaris:2006zy}%
  \BibitemOpen
  \bibfield  {author} {\bibinfo {author} {\bibfnamefont {C.}~\bibnamefont
  {Kouvaris}}, \bibinfo {author} {\bibfnamefont {J.-W.}\ \bibnamefont {Qiu}},
  \bibinfo {author} {\bibfnamefont {W.}~\bibnamefont {Vogelsang}}, \ and\
  \bibinfo {author} {\bibfnamefont {F.}~\bibnamefont {Yuan}},\ }\href {\doibase
  10.1103/PhysRevD.74.114013} {\bibfield  {journal} {\bibinfo  {journal}
  {Phys.~Rev.}\ }\textbf {\bibinfo {volume} {D74}},\ \bibinfo {pages} {114013}
  (\bibinfo {year} {2006})},\ \Eprint {http://arxiv.org/abs/hep-ph/0609238}
  {hep-ph/0609238} \BibitemShut {NoStop}%
\bibitem [{\citenamefont {Eguchi}\ \emph {et~al.}(2007)\citenamefont {Eguchi},
  \citenamefont {Koike},\ and\ \citenamefont {Tanaka}}]{Eguchi:2006mc}%
  \BibitemOpen
  \bibfield  {author} {\bibinfo {author} {\bibfnamefont {H.}~\bibnamefont
  {Eguchi}}, \bibinfo {author} {\bibfnamefont {Y.}~\bibnamefont {Koike}}, \
  and\ \bibinfo {author} {\bibfnamefont {K.}~\bibnamefont {Tanaka}},\
  }\href@noop {} {\bibfield  {journal} {\bibinfo  {journal} {Nucl.~Phys.}\
  }\textbf {\bibinfo {volume} {B763}},\ \bibinfo {pages} {198} (\bibinfo {year}
  {2007})},\ \Eprint {http://arxiv.org/abs/hep-ph/0610314} {hep-ph/0610314}
  \BibitemShut {NoStop}%
\bibitem [{\citenamefont {Koike}\ and\ \citenamefont
  {Tomita}(2009)}]{Koike:2009ge}%
  \BibitemOpen
  \bibfield  {author} {\bibinfo {author} {\bibfnamefont {Y.}~\bibnamefont
  {Koike}}\ and\ \bibinfo {author} {\bibfnamefont {T.}~\bibnamefont {Tomita}},\
  }\href {\doibase 10.1016/j.physletb.2009.04.017} {\bibfield  {journal}
  {\bibinfo  {journal} {Phys.~Lett.}\ }\textbf {\bibinfo {volume} {B675}},\
  \bibinfo {pages} {181} (\bibinfo {year} {2009})},\ \Eprint
  {http://arxiv.org/abs/0903.1923} {arXiv:0903.1923 [hep-ph]} \BibitemShut
  {NoStop}%
\bibitem [{\citenamefont {Kang}\ \emph {et~al.}(2011)\citenamefont {Kang},
  \citenamefont {Qiu}, \citenamefont {Vogelsang},\ and\ \citenamefont
  {Yuan}}]{Kang:2011hk}%
  \BibitemOpen
  \bibfield  {author} {\bibinfo {author} {\bibfnamefont {Z.-B.}\ \bibnamefont
  {Kang}}, \bibinfo {author} {\bibfnamefont {J.-W.}\ \bibnamefont {Qiu}},
  \bibinfo {author} {\bibfnamefont {W.}~\bibnamefont {Vogelsang}}, \ and\
  \bibinfo {author} {\bibfnamefont {F.}~\bibnamefont {Yuan}},\ }\href {\doibase
  10.1103/PhysRevD.83.094001} {\bibfield  {journal} {\bibinfo  {journal}
  {Phys.~Rev.}\ }\textbf {\bibinfo {volume} {D83}},\ \bibinfo {pages} {094001}
  (\bibinfo {year} {2011})},\ \Eprint {http://arxiv.org/abs/1103.1591}
  {arXiv:1103.1591 [hep-ph]} \BibitemShut {NoStop}%
\bibitem [{\citenamefont {Metz}\ and\ \citenamefont
  {Pitonyak}(2013)}]{Metz:2012ct}%
  \BibitemOpen
  \bibfield  {author} {\bibinfo {author} {\bibfnamefont {A.}~\bibnamefont
  {Metz}}\ and\ \bibinfo {author} {\bibfnamefont {D.}~\bibnamefont
  {Pitonyak}},\ }\href {\doibase 10.1016/j.physletb.2013.05.043} {\bibfield
  {journal} {\bibinfo  {journal} {Phys.~Lett.}\ }\textbf {\bibinfo {volume}
  {B723}},\ \bibinfo {pages} {365} (\bibinfo {year} {2013})},\ \Eprint
  {http://arxiv.org/abs/1212.5037} {arXiv:1212.5037 [hep-ph]} \BibitemShut
  {NoStop}%
\bibitem [{\citenamefont {Beppu}\ \emph {et~al.}(2014)\citenamefont {Beppu},
  \citenamefont {Kanazawa}, \citenamefont {Koike},\ and\ \citenamefont
  {Yoshida}}]{Beppu:2013uda}%
  \BibitemOpen
  \bibfield  {author} {\bibinfo {author} {\bibfnamefont {H.}~\bibnamefont
  {Beppu}}, \bibinfo {author} {\bibfnamefont {K.}~\bibnamefont {Kanazawa}},
  \bibinfo {author} {\bibfnamefont {Y.}~\bibnamefont {Koike}}, \ and\ \bibinfo
  {author} {\bibfnamefont {S.}~\bibnamefont {Yoshida}},\ }\href {\doibase
  10.1103/PhysRevD.89.034029} {\bibfield  {journal} {\bibinfo  {journal}
  {Phys.~Rev.}\ }\textbf {\bibinfo {volume} {D89}},\ \bibinfo {pages} {034029}
  (\bibinfo {year} {2014})},\ \Eprint {http://arxiv.org/abs/1312.6862}
  {arXiv:1312.6862 [hep-ph]} \BibitemShut {NoStop}%
\bibitem [{\citenamefont {Ji}\ \emph {et~al.}(2004)\citenamefont {Ji},
  \citenamefont {Ma},\ and\ \citenamefont {Yuan}}]{Ji:2004xq}%
  \BibitemOpen
  \bibfield  {author} {\bibinfo {author} {\bibfnamefont {X.-d.}\ \bibnamefont
  {Ji}}, \bibinfo {author} {\bibfnamefont {J.-P.}\ \bibnamefont {Ma}}, \ and\
  \bibinfo {author} {\bibfnamefont {F.}~\bibnamefont {Yuan}},\ }\href@noop {}
  {\bibfield  {journal} {\bibinfo  {journal} {Phys. Lett.}\ }\textbf {\bibinfo
  {volume} {B597}},\ \bibinfo {pages} {299} (\bibinfo {year} {2004})},\ \Eprint
  {http://arxiv.org/abs/hep-ph/0405085} {hep-ph/0405085} \BibitemShut {NoStop}%
\bibitem [{\citenamefont {Collins}(2011)}]{Collins:2011zzd}%
  \BibitemOpen
  \bibfield  {author} {\bibinfo {author} {\bibfnamefont {J.}~\bibnamefont
  {Collins}},\ }\href@noop {} {\bibfield  {journal} {\bibinfo  {journal} {Camb.
  Monogr. Part. Phys. Nucl. Phys. Cosmol.}\ }\textbf {\bibinfo {volume} {32}},\
  \bibinfo {pages} {1} (\bibinfo {year} {2011})}\BibitemShut {NoStop}%
\bibitem [{\citenamefont {Kotzinian}(1995)}]{Kotzinian:1994dv}%
  \BibitemOpen
  \bibfield  {author} {\bibinfo {author} {\bibfnamefont {A.}~\bibnamefont
  {Kotzinian}},\ }\href {\doibase 10.1016/0550-3213(95)00098-D} {\bibfield
  {journal} {\bibinfo  {journal} {Nucl. Phys.}\ }\textbf {\bibinfo {volume}
  {B441}},\ \bibinfo {pages} {234} (\bibinfo {year} {1995})},\ \Eprint
  {http://arxiv.org/abs/hep-ph/9412283} {arXiv:hep-ph/9412283} \BibitemShut
  {NoStop}%
\bibitem [{\citenamefont {Mulders}\ and\ \citenamefont
  {Tangerman}(1996)}]{Mulders:1995dh}%
  \BibitemOpen
  \bibfield  {author} {\bibinfo {author} {\bibfnamefont {P.~J.}\ \bibnamefont
  {Mulders}}\ and\ \bibinfo {author} {\bibfnamefont {R.~D.}\ \bibnamefont
  {Tangerman}},\ }\href {\doibase 10.1016/0550-3213(95)00632-X} {\bibfield
  {journal} {\bibinfo  {journal} {Nucl. Phys.}\ }\textbf {\bibinfo {volume}
  {B461}},\ \bibinfo {pages} {197} (\bibinfo {year} {1996})},\ \Eprint
  {http://arxiv.org/abs/hep-ph/9510301} {arXiv:hep-ph/9510301} \BibitemShut
  {NoStop}%
\bibitem [{\citenamefont {Boer}\ \emph {et~al.}(1997)\citenamefont {Boer},
  \citenamefont {Jakob},\ and\ \citenamefont {Mulders}}]{Boer:1997mf}%
  \BibitemOpen
  \bibfield  {author} {\bibinfo {author} {\bibfnamefont {D.}~\bibnamefont
  {Boer}}, \bibinfo {author} {\bibfnamefont {R.}~\bibnamefont {Jakob}}, \ and\
  \bibinfo {author} {\bibfnamefont {P.~J.}\ \bibnamefont {Mulders}},\ }\href
  {\doibase 10.1016/S0550-3213(97)00456-2} {\bibfield  {journal} {\bibinfo
  {journal} {Nucl. Phys.}\ }\textbf {\bibinfo {volume} {B504}},\ \bibinfo
  {pages} {345} (\bibinfo {year} {1997})},\ \Eprint
  {http://arxiv.org/abs/hep-ph/9702281} {arXiv:hep-ph/9702281 [hep-ph]}
  \BibitemShut {NoStop}%
\bibitem [{\citenamefont {Bacchetta}\ \emph {et~al.}(2007)\citenamefont
  {Bacchetta}, \citenamefont {Diehl}, \citenamefont {Goeke}, \citenamefont
  {Metz}, \citenamefont {Mulders} \emph {et~al.}}]{Bacchetta:2006tn}%
  \BibitemOpen
  \bibfield  {author} {\bibinfo {author} {\bibfnamefont {A.}~\bibnamefont
  {Bacchetta}}, \bibinfo {author} {\bibfnamefont {M.}~\bibnamefont {Diehl}},
  \bibinfo {author} {\bibfnamefont {K.}~\bibnamefont {Goeke}}, \bibinfo
  {author} {\bibfnamefont {A.}~\bibnamefont {Metz}}, \bibinfo {author}
  {\bibfnamefont {P.~J.}\ \bibnamefont {Mulders}},  \emph {et~al.},\ }\href
  {\doibase 10.1088/1126-6708/2007/02/093} {\bibfield  {journal} {\bibinfo
  {journal} {JHEP}\ }\textbf {\bibinfo {volume} {0702}},\ \bibinfo {pages}
  {093} (\bibinfo {year} {2007})},\ \Eprint
  {http://arxiv.org/abs/hep-ph/0611265} {arXiv:hep-ph/0611265 [hep-ph]}
  \BibitemShut {NoStop}%
\bibitem [{\citenamefont {Arnold}\ \emph {et~al.}(2009)\citenamefont {Arnold},
  \citenamefont {Metz},\ and\ \citenamefont {Schlegel}}]{Arnold:2008kf}%
  \BibitemOpen
  \bibfield  {author} {\bibinfo {author} {\bibfnamefont {S.}~\bibnamefont
  {Arnold}}, \bibinfo {author} {\bibfnamefont {A.}~\bibnamefont {Metz}}, \ and\
  \bibinfo {author} {\bibfnamefont {M.}~\bibnamefont {Schlegel}},\ }\href
  {\doibase 10.1103/PhysRevD.79.034005} {\bibfield  {journal} {\bibinfo
  {journal} {Phys. Rev.}\ }\textbf {\bibinfo {volume} {D79}},\ \bibinfo {pages}
  {034005} (\bibinfo {year} {2009})},\ \Eprint {http://arxiv.org/abs/0809.2262}
  {arXiv:0809.2262 [hep-ph]} \BibitemShut {NoStop}%
\bibitem [{\citenamefont {Pitonyak}\ \emph {et~al.}(2014)\citenamefont
  {Pitonyak}, \citenamefont {Schlegel},\ and\ \citenamefont
  {Metz}}]{Pitonyak:2013dsu}%
  \BibitemOpen
  \bibfield  {author} {\bibinfo {author} {\bibfnamefont {D.}~\bibnamefont
  {Pitonyak}}, \bibinfo {author} {\bibfnamefont {M.}~\bibnamefont {Schlegel}},
  \ and\ \bibinfo {author} {\bibfnamefont {A.}~\bibnamefont {Metz}},\ }\href
  {\doibase 10.1103/PhysRevD.89.054032} {\bibfield  {journal} {\bibinfo
  {journal} {Phys. Rev.}\ }\textbf {\bibinfo {volume} {D89}},\ \bibinfo {pages}
  {054032} (\bibinfo {year} {2014})},\ \Eprint {http://arxiv.org/abs/1310.6240}
  {arXiv:1310.6240 [hep-ph]} \BibitemShut {NoStop}%
\bibitem [{\citenamefont {Ji}\ \emph {et~al.}(2006{\natexlab{a}})\citenamefont
  {Ji}, \citenamefont {Qiu}, \citenamefont {Vogelsang},\ and\ \citenamefont
  {Yuan}}]{Ji:2006ub}%
  \BibitemOpen
  \bibfield  {author} {\bibinfo {author} {\bibfnamefont {X.}~\bibnamefont
  {Ji}}, \bibinfo {author} {\bibfnamefont {J.-W.}\ \bibnamefont {Qiu}},
  \bibinfo {author} {\bibfnamefont {W.}~\bibnamefont {Vogelsang}}, \ and\
  \bibinfo {author} {\bibfnamefont {F.}~\bibnamefont {Yuan}},\ }\href {\doibase
  10.1103/PhysRevLett.97.082002} {\bibfield  {journal} {\bibinfo  {journal}
  {Phys. Rev. Lett.}\ }\textbf {\bibinfo {volume} {97}},\ \bibinfo {pages}
  {082002} (\bibinfo {year} {2006}{\natexlab{a}})},\ \Eprint
  {http://arxiv.org/abs/hep-ph/0602239} {arXiv:hep-ph/0602239} \BibitemShut
  {NoStop}%
\bibitem [{\citenamefont {Ji}\ \emph {et~al.}(2006{\natexlab{b}})\citenamefont
  {Ji}, \citenamefont {Qiu}, \citenamefont {Vogelsang},\ and\ \citenamefont
  {Yuan}}]{Ji:2006br}%
  \BibitemOpen
  \bibfield  {author} {\bibinfo {author} {\bibfnamefont {X.}~\bibnamefont
  {Ji}}, \bibinfo {author} {\bibfnamefont {J.-W.}\ \bibnamefont {Qiu}},
  \bibinfo {author} {\bibfnamefont {W.}~\bibnamefont {Vogelsang}}, \ and\
  \bibinfo {author} {\bibfnamefont {F.}~\bibnamefont {Yuan}},\ }\href {\doibase
  10.1016/j.physletb.2006.05.044} {\bibfield  {journal} {\bibinfo  {journal}
  {Phys. Lett.}\ }\textbf {\bibinfo {volume} {B638}},\ \bibinfo {pages} {178}
  (\bibinfo {year} {2006}{\natexlab{b}})},\ \Eprint
  {http://arxiv.org/abs/hep-ph/0604128} {arXiv:hep-ph/0604128} \BibitemShut
  {NoStop}%
\bibitem [{\citenamefont {Koike}\ \emph {et~al.}(2008)\citenamefont {Koike},
  \citenamefont {Vogelsang},\ and\ \citenamefont {Yuan}}]{Koike:2007dg}%
  \BibitemOpen
  \bibfield  {author} {\bibinfo {author} {\bibfnamefont {Y.}~\bibnamefont
  {Koike}}, \bibinfo {author} {\bibfnamefont {W.}~\bibnamefont {Vogelsang}}, \
  and\ \bibinfo {author} {\bibfnamefont {F.}~\bibnamefont {Yuan}},\ }\href
  {\doibase 10.1016/j.physletb.2007.11.096} {\bibfield  {journal} {\bibinfo
  {journal} {Phys.Lett.}\ }\textbf {\bibinfo {volume} {B659}},\ \bibinfo
  {pages} {878} (\bibinfo {year} {2008})},\ \Eprint
  {http://arxiv.org/abs/0711.0636} {arXiv:0711.0636 [hep-ph]} \BibitemShut
  {NoStop}%
\bibitem [{\citenamefont {Zhou}\ \emph {et~al.}(2008)\citenamefont {Zhou},
  \citenamefont {Yuan},\ and\ \citenamefont {Liang}}]{Zhou:2008fb}%
  \BibitemOpen
  \bibfield  {author} {\bibinfo {author} {\bibfnamefont {J.}~\bibnamefont
  {Zhou}}, \bibinfo {author} {\bibfnamefont {F.}~\bibnamefont {Yuan}}, \ and\
  \bibinfo {author} {\bibfnamefont {Z.-T.}\ \bibnamefont {Liang}},\ }\href
  {\doibase 10.1103/PhysRevD.78.114008} {\bibfield  {journal} {\bibinfo
  {journal} {Phys. Rev.}\ }\textbf {\bibinfo {volume} {D78}},\ \bibinfo {pages}
  {114008} (\bibinfo {year} {2008})},\ \Eprint {http://arxiv.org/abs/0808.3629}
  {arXiv:0808.3629 [hep-ph]} \BibitemShut {NoStop}%
\bibitem [{\citenamefont {Yuan}\ and\ \citenamefont
  {Zhou}(2009)}]{Yuan:2009dw}%
  \BibitemOpen
  \bibfield  {author} {\bibinfo {author} {\bibfnamefont {F.}~\bibnamefont
  {Yuan}}\ and\ \bibinfo {author} {\bibfnamefont {J.}~\bibnamefont {Zhou}},\
  }\href {\doibase 10.1103/PhysRevLett.103.052001} {\bibfield  {journal}
  {\bibinfo  {journal} {Phys.~Rev.~Lett.}\ }\textbf {\bibinfo {volume} {103}},\
  \bibinfo {pages} {052001} (\bibinfo {year} {2009})},\ \Eprint
  {http://arxiv.org/abs/0903.4680} {arXiv:0903.4680 [hep-ph]} \BibitemShut
  {NoStop}%
\bibitem [{\citenamefont {Zhou}\ \emph {et~al.}(2010)\citenamefont {Zhou},
  \citenamefont {Yuan},\ and\ \citenamefont {Liang}}]{Zhou:2009jm}%
  \BibitemOpen
  \bibfield  {author} {\bibinfo {author} {\bibfnamefont {J.}~\bibnamefont
  {Zhou}}, \bibinfo {author} {\bibfnamefont {F.}~\bibnamefont {Yuan}}, \ and\
  \bibinfo {author} {\bibfnamefont {Z.-T.}\ \bibnamefont {Liang}},\ }\href
  {\doibase 10.1103/PhysRevD.81.054008} {\bibfield  {journal} {\bibinfo
  {journal} {Phys. Rev.}\ }\textbf {\bibinfo {volume} {D81}},\ \bibinfo {pages}
  {054008} (\bibinfo {year} {2010})},\ \Eprint {http://arxiv.org/abs/0909.2238}
  {arXiv:0909.2238 [hep-ph]} \BibitemShut {NoStop}%
\bibitem [{\citenamefont {Aschenauer}\ \emph {et~al.}(2015)\citenamefont
  {Aschenauer} \emph {et~al.}}]{Aschenauer:2015eha}%
  \BibitemOpen
  \bibfield  {author} {\bibinfo {author} {\bibfnamefont {E.-C.}\ \bibnamefont
  {Aschenauer}} \emph {et~al.},\ }\href@noop {} {\  (\bibinfo {year} {2015})},\
  \Eprint {http://arxiv.org/abs/1501.01220} {arXiv:1501.01220 [nucl-ex]}
  \BibitemShut {NoStop}%
\bibitem [{\citenamefont {Gautheron}\ \emph {et~al.}(2010)\citenamefont
  {Gautheron} \emph {et~al.}}]{Gautheron:2010wva}%
  \BibitemOpen
  \bibfield  {author} {\bibinfo {author} {\bibfnamefont {F.}~\bibnamefont
  {Gautheron}} \emph {et~al.} (\bibinfo {collaboration} {COMPASS}),\
  }\href@noop {} {\bibfield  {journal} {\bibinfo  {journal}
  {CERN-SPSC-2010-014}\ } (\bibinfo {year} {2010})}\BibitemShut {NoStop}%
\bibitem [{\citenamefont {Bradamante}(2018)}]{Bradamante:2018ick}%
  \BibitemOpen
  \bibfield  {author} {\bibinfo {author} {\bibfnamefont {F.}~\bibnamefont
  {Bradamante}} (\bibinfo {collaboration} {COMPASS}),\ }\bibfield  {booktitle}
  {\emph {\bibinfo {booktitle} {{Proceedings, 23rd International Symposium on
  Spin Physics (SPIN 2018): Ferrara, Italy, September 10-14, 2018}}},\ }\href
  {\doibase 10.22323/1.346.0045} {\bibfield  {journal} {\bibinfo  {journal}
  {PoS}\ }\textbf {\bibinfo {volume} {SPIN2018}},\ \bibinfo {pages} {045}
  (\bibinfo {year} {2018})},\ \Eprint {http://arxiv.org/abs/1812.07281}
  {arXiv:1812.07281 [hep-ex]} \BibitemShut {NoStop}%
\bibitem [{\citenamefont {Dudek}\ \emph {et~al.}(2012)\citenamefont {Dudek},
  \citenamefont {Ent}, \citenamefont {Essig}, \citenamefont {Kumar},
  \citenamefont {Meyer} \emph {et~al.}}]{Dudek:2012vr}%
  \BibitemOpen
  \bibfield  {author} {\bibinfo {author} {\bibfnamefont {J.}~\bibnamefont
  {Dudek}}, \bibinfo {author} {\bibfnamefont {R.}~\bibnamefont {Ent}}, \bibinfo
  {author} {\bibfnamefont {R.}~\bibnamefont {Essig}}, \bibinfo {author}
  {\bibfnamefont {K.}~\bibnamefont {Kumar}}, \bibinfo {author} {\bibfnamefont
  {C.}~\bibnamefont {Meyer}},  \emph {et~al.},\ }\href {\doibase
  10.1140/epja/i2012-12187-1} {\bibfield  {journal} {\bibinfo  {journal}
  {Eur.~Phys.~J.}\ }\textbf {\bibinfo {volume} {A48}},\ \bibinfo {pages} {187}
  (\bibinfo {year} {2012})},\ \Eprint {http://arxiv.org/abs/1208.1244}
  {arXiv:1208.1244 [hep-ex]} \BibitemShut {NoStop}%
\bibitem [{\citenamefont {Altmannshofer}\ \emph {et~al.}(2019)\citenamefont
  {Altmannshofer} \emph {et~al.}}]{Kou:2018nap}%
  \BibitemOpen
  \bibfield  {author} {\bibinfo {author} {\bibfnamefont {W.}~\bibnamefont
  {Altmannshofer}} \emph {et~al.} (\bibinfo {collaboration} {Belle-II}),\
  }\href {\doibase 10.1093/ptep/ptz106} {\bibfield  {journal} {\bibinfo
  {journal} {PTEP}\ }\textbf {\bibinfo {volume} {2019}},\ \bibinfo {pages}
  {123C01} (\bibinfo {year} {2019})},\ \Eprint
  {http://arxiv.org/abs/1808.10567} {arXiv:1808.10567 [hep-ex]} \BibitemShut
  {NoStop}%
\bibitem [{\citenamefont {Boer}\ \emph
  {et~al.}(2011{\natexlab{a}})\citenamefont {Boer}, \citenamefont {Diehl},
  \citenamefont {Milner}, \citenamefont {Venugopalan}, \citenamefont
  {Vogelsang} \emph {et~al.}}]{Boer:2011fh}%
  \BibitemOpen
  \bibfield  {author} {\bibinfo {author} {\bibfnamefont {D.}~\bibnamefont
  {Boer}}, \bibinfo {author} {\bibfnamefont {M.}~\bibnamefont {Diehl}},
  \bibinfo {author} {\bibfnamefont {R.}~\bibnamefont {Milner}}, \bibinfo
  {author} {\bibfnamefont {R.}~\bibnamefont {Venugopalan}}, \bibinfo {author}
  {\bibfnamefont {W.}~\bibnamefont {Vogelsang}},  \emph {et~al.},\ }\href@noop
  {} {\  (\bibinfo {year} {2011}{\natexlab{a}})},\ \Eprint
  {http://arxiv.org/abs/1108.1713} {arXiv:1108.1713 [nucl-th]} \BibitemShut
  {NoStop}%
\bibitem [{\citenamefont {Accardi}\ \emph
  {et~al.}(2016{\natexlab{a}})\citenamefont {Accardi} \emph
  {et~al.}}]{Accardi:2012qut}%
  \BibitemOpen
  \bibfield  {author} {\bibinfo {author} {\bibfnamefont {A.}~\bibnamefont
  {Accardi}} \emph {et~al.},\ }\href {\doibase 10.1140/epja/i2016-16268-9}
  {\bibfield  {journal} {\bibinfo  {journal} {Eur. Phys. J.}\ }\textbf
  {\bibinfo {volume} {A52}},\ \bibinfo {pages} {268} (\bibinfo {year}
  {2016}{\natexlab{a}})},\ \Eprint {http://arxiv.org/abs/1212.1701}
  {arXiv:1212.1701 [nucl-ex]} \BibitemShut {NoStop}%
\bibitem [{\citenamefont {Collins}\ and\ \citenamefont
  {Soper}(1982)}]{Collins:1981uw}%
  \BibitemOpen
  \bibfield  {author} {\bibinfo {author} {\bibfnamefont {J.~C.}\ \bibnamefont
  {Collins}}\ and\ \bibinfo {author} {\bibfnamefont {D.~E.}\ \bibnamefont
  {Soper}},\ }\href@noop {} {\bibfield  {journal} {\bibinfo  {journal} {Nucl.
  Phys.}\ }\textbf {\bibinfo {volume} {B194}},\ \bibinfo {pages} {445}
  (\bibinfo {year} {1982})}\BibitemShut {NoStop}%
\bibitem [{\citenamefont {Boer}\ \emph
  {et~al.}(2011{\natexlab{b}})\citenamefont {Boer}, \citenamefont {Gamberg},
  \citenamefont {Musch},\ and\ \citenamefont {Prokudin}}]{Boer:2011xd}%
  \BibitemOpen
  \bibfield  {author} {\bibinfo {author} {\bibfnamefont {D.}~\bibnamefont
  {Boer}}, \bibinfo {author} {\bibfnamefont {L.}~\bibnamefont {Gamberg}},
  \bibinfo {author} {\bibfnamefont {B.}~\bibnamefont {Musch}}, \ and\ \bibinfo
  {author} {\bibfnamefont {A.}~\bibnamefont {Prokudin}},\ }\href {\doibase
  10.1007/JHEP10(2011)021} {\bibfield  {journal} {\bibinfo  {journal} {JHEP}\
  }\textbf {\bibinfo {volume} {1110}},\ \bibinfo {pages} {021} (\bibinfo {year}
  {2011}{\natexlab{b}})},\ \Eprint {http://arxiv.org/abs/1107.5294}
  {arXiv:1107.5294 [hep-ph]} \BibitemShut {NoStop}%
\bibitem [{\citenamefont {Aybat}\ \emph {et~al.}(2012)\citenamefont {Aybat},
  \citenamefont {Collins}, \citenamefont {Qiu},\ and\ \citenamefont
  {Rogers}}]{Aybat:2011ge}%
  \BibitemOpen
  \bibfield  {author} {\bibinfo {author} {\bibfnamefont {S.~M.}\ \bibnamefont
  {Aybat}}, \bibinfo {author} {\bibfnamefont {J.~C.}\ \bibnamefont {Collins}},
  \bibinfo {author} {\bibfnamefont {J.-W.}\ \bibnamefont {Qiu}}, \ and\
  \bibinfo {author} {\bibfnamefont {T.~C.}\ \bibnamefont {Rogers}},\
  }\href@noop {} {\bibfield  {journal} {\bibinfo  {journal} {Phys.~Rev.}\
  }\textbf {\bibinfo {volume} {D85}},\ \bibinfo {pages} {034043} (\bibinfo
  {year} {2012})},\ \Eprint {http://arxiv.org/abs/1110.6428} {arXiv:1110.6428
  [hep-ph]} \BibitemShut {NoStop}%
\bibitem [{\citenamefont {Kanazawa}\ \emph {et~al.}(2016)\citenamefont
  {Kanazawa}, \citenamefont {Koike}, \citenamefont {Metz}, \citenamefont
  {Pitonyak},\ and\ \citenamefont {Schlegel}}]{Kanazawa:2015ajw}%
  \BibitemOpen
  \bibfield  {author} {\bibinfo {author} {\bibfnamefont {K.}~\bibnamefont
  {Kanazawa}}, \bibinfo {author} {\bibfnamefont {Y.}~\bibnamefont {Koike}},
  \bibinfo {author} {\bibfnamefont {A.}~\bibnamefont {Metz}}, \bibinfo {author}
  {\bibfnamefont {D.}~\bibnamefont {Pitonyak}}, \ and\ \bibinfo {author}
  {\bibfnamefont {M.}~\bibnamefont {Schlegel}},\ }\href {\doibase
  10.1103/PhysRevD.93.054024} {\bibfield  {journal} {\bibinfo  {journal} {Phys.
  Rev.}\ }\textbf {\bibinfo {volume} {D93}},\ \bibinfo {pages} {054024}
  (\bibinfo {year} {2016})},\ \Eprint {http://arxiv.org/abs/1512.07233}
  {arXiv:1512.07233 [hep-ph]} \BibitemShut {NoStop}%
\bibitem [{\citenamefont {Gamberg}\ \emph {et~al.}(2018)\citenamefont
  {Gamberg}, \citenamefont {Metz}, \citenamefont {Pitonyak},\ and\
  \citenamefont {Prokudin}}]{Gamberg:2017jha}%
  \BibitemOpen
  \bibfield  {author} {\bibinfo {author} {\bibfnamefont {L.}~\bibnamefont
  {Gamberg}}, \bibinfo {author} {\bibfnamefont {A.}~\bibnamefont {Metz}},
  \bibinfo {author} {\bibfnamefont {D.}~\bibnamefont {Pitonyak}}, \ and\
  \bibinfo {author} {\bibfnamefont {A.}~\bibnamefont {Prokudin}},\ }\href
  {\doibase 10.1016/j.physletb.2018.03.024} {\bibfield  {journal} {\bibinfo
  {journal} {Phys. Lett.}\ }\textbf {\bibinfo {volume} {B781}},\ \bibinfo
  {pages} {443} (\bibinfo {year} {2018})},\ \Eprint
  {http://arxiv.org/abs/1712.08116} {arXiv:1712.08116 [hep-ph]} \BibitemShut
  {NoStop}%
\bibitem [{\citenamefont {Scimemi}\ \emph {et~al.}(2019)\citenamefont
  {Scimemi}, \citenamefont {Tarasov},\ and\ \citenamefont
  {Vladimirov}}]{Scimemi:2019gge}%
  \BibitemOpen
  \bibfield  {author} {\bibinfo {author} {\bibfnamefont {I.}~\bibnamefont
  {Scimemi}}, \bibinfo {author} {\bibfnamefont {A.}~\bibnamefont {Tarasov}}, \
  and\ \bibinfo {author} {\bibfnamefont {A.}~\bibnamefont {Vladimirov}},\
  }\href {\doibase 10.1007/JHEP05(2019)125} {\bibfield  {journal} {\bibinfo
  {journal} {JHEP}\ }\textbf {\bibinfo {volume} {05}},\ \bibinfo {pages} {125}
  (\bibinfo {year} {2019})},\ \Eprint {http://arxiv.org/abs/1901.04519}
  {arXiv:1901.04519 [hep-ph]} \BibitemShut {NoStop}%
\bibitem [{\citenamefont {Boer}\ \emph {et~al.}(2003)\citenamefont {Boer},
  \citenamefont {Mulders},\ and\ \citenamefont {Pijlman}}]{Boer:2003cm}%
  \BibitemOpen
  \bibfield  {author} {\bibinfo {author} {\bibfnamefont {D.}~\bibnamefont
  {Boer}}, \bibinfo {author} {\bibfnamefont {P.~J.}\ \bibnamefont {Mulders}}, \
  and\ \bibinfo {author} {\bibfnamefont {F.}~\bibnamefont {Pijlman}},\
  }\href@noop {} {\bibfield  {journal} {\bibinfo  {journal} {Nucl.~Phys.}\
  }\textbf {\bibinfo {volume} {B667}},\ \bibinfo {pages} {201} (\bibinfo {year}
  {2003})},\ \Eprint {http://arxiv.org/abs/hep-ph/0303034} {hep-ph/0303034}
  \BibitemShut {NoStop}%
\bibitem [{\citenamefont {Sivers}(1990)}]{Sivers:1989cc}%
  \BibitemOpen
  \bibfield  {author} {\bibinfo {author} {\bibfnamefont {D.~W.}\ \bibnamefont
  {Sivers}},\ }\href@noop {} {\bibfield  {journal} {\bibinfo  {journal}
  {Phys.~Rev.}\ }\textbf {\bibinfo {volume} {D41}},\ \bibinfo {pages} {83}
  (\bibinfo {year} {1990})}\BibitemShut {NoStop}%
\bibitem [{\citenamefont {Sivers}(1991)}]{Sivers:1990fh}%
  \BibitemOpen
  \bibfield  {author} {\bibinfo {author} {\bibfnamefont {D.~W.}\ \bibnamefont
  {Sivers}},\ }\href@noop {} {\bibfield  {journal} {\bibinfo  {journal}
  {Phys.~Rev.}\ }\textbf {\bibinfo {volume} {D43}},\ \bibinfo {pages} {261}
  (\bibinfo {year} {1991})}\BibitemShut {NoStop}%
\bibitem [{\citenamefont {Qiu}\ \emph {et~al.}(2020)\citenamefont {Qiu},
  \citenamefont {Rogers},\ and\ \citenamefont {Wang}}]{Qiu:2020oqr}%
  \BibitemOpen
  \bibfield  {author} {\bibinfo {author} {\bibfnamefont {J.-W.}\ \bibnamefont
  {Qiu}}, \bibinfo {author} {\bibfnamefont {T.~C.}\ \bibnamefont {Rogers}}, \
  and\ \bibinfo {author} {\bibfnamefont {B.}~\bibnamefont {Wang}},\ }\href@noop
  {} {\  (\bibinfo {year} {2020})},\ \Eprint {http://arxiv.org/abs/2004.13193}
  {arXiv:2004.13193 [hep-ph]} \BibitemShut {NoStop}%
\bibitem [{\citenamefont {Airapetian.}\ \emph {et~al.}(2009)\citenamefont
  {Airapetian.} \emph {et~al.}}]{Airapetian:2009ae}%
  \BibitemOpen
  \bibfield  {author} {\bibinfo {author} {\bibfnamefont {A.}~\bibnamefont
  {Airapetian.}} \emph {et~al.} (\bibinfo {collaboration} {HERMES}),\
  }\href@noop {} {\bibfield  {journal} {\bibinfo  {journal} {Phys. Rev. Lett.}\
  }\textbf {\bibinfo {volume} {103}},\ \bibinfo {pages} {152002} (\bibinfo
  {year} {2009})},\ \Eprint {http://arxiv.org/abs/0906.3918} {arXiv:0906.3918
  [hep-ex]} \BibitemShut {NoStop}%
\bibitem [{\citenamefont {Alekseev}\ \emph {et~al.}(2009)\citenamefont
  {Alekseev} \emph {et~al.}}]{Alekseev:2008aa}%
  \BibitemOpen
  \bibfield  {author} {\bibinfo {author} {\bibfnamefont {M.}~\bibnamefont
  {Alekseev}} \emph {et~al.} (\bibinfo {collaboration} {COMPASS}),\ }\href
  {\doibase 10.1016/j.physletb.2009.01.060} {\bibfield  {journal} {\bibinfo
  {journal} {Phys.Lett.}\ }\textbf {\bibinfo {volume} {B673}},\ \bibinfo
  {pages} {127} (\bibinfo {year} {2009})},\ \Eprint
  {http://arxiv.org/abs/0802.2160} {arXiv:0802.2160 [hep-ex]} \BibitemShut
  {NoStop}%
\bibitem [{\citenamefont {Qian}\ \emph {et~al.}(2011)\citenamefont {Qian} \emph
  {et~al.}}]{Qian:2011py}%
  \BibitemOpen
  \bibfield  {author} {\bibinfo {author} {\bibfnamefont {X.}~\bibnamefont
  {Qian}} \emph {et~al.} (\bibinfo {collaboration} {The Jefferson Lab Hall
  A}),\ }\href {\doibase 10.1103/PhysRevLett.107.072003} {\bibfield  {journal}
  {\bibinfo  {journal} {Phys.Rev.Lett.}\ }\textbf {\bibinfo {volume} {107}},\
  \bibinfo {pages} {072003} (\bibinfo {year} {2011})},\ \Eprint
  {http://arxiv.org/abs/1106.0363} {arXiv:1106.0363 [nucl-ex]} \BibitemShut
  {NoStop}%
\bibitem [{\citenamefont {Adolph}\ \emph {et~al.}(2015)\citenamefont {Adolph}
  \emph {et~al.}}]{Adolph:2014zba}%
  \BibitemOpen
  \bibfield  {author} {\bibinfo {author} {\bibfnamefont {C.}~\bibnamefont
  {Adolph}} \emph {et~al.} (\bibinfo {collaboration} {COMPASS}),\ }\href
  {\doibase 10.1016/j.physletb.2015.03.056} {\bibfield  {journal} {\bibinfo
  {journal} {Phys. Lett.}\ }\textbf {\bibinfo {volume} {B744}},\ \bibinfo
  {pages} {250} (\bibinfo {year} {2015})},\ \Eprint
  {http://arxiv.org/abs/1408.4405} {arXiv:1408.4405 [hep-ex]} \BibitemShut
  {NoStop}%
\bibitem [{\citenamefont {Zhao}\ \emph {et~al.}(2014)\citenamefont {Zhao} \emph
  {et~al.}}]{Zhao:2014qvx}%
  \BibitemOpen
  \bibfield  {author} {\bibinfo {author} {\bibfnamefont {Y.~X.}\ \bibnamefont
  {Zhao}} \emph {et~al.} (\bibinfo {collaboration} {Jefferson Lab Hall A}),\
  }\href {\doibase 10.1103/PhysRevC.90.055201} {\bibfield  {journal} {\bibinfo
  {journal} {Phys. Rev.}\ }\textbf {\bibinfo {volume} {C90}},\ \bibinfo {pages}
  {055201} (\bibinfo {year} {2014})},\ \Eprint {http://arxiv.org/abs/1404.7204}
  {arXiv:1404.7204 [nucl-ex]} \BibitemShut {NoStop}%
\bibitem [{\citenamefont {Adolph}\ \emph {et~al.}(2017)\citenamefont {Adolph}
  \emph {et~al.}}]{Adolph:2016dvl}%
  \BibitemOpen
  \bibfield  {author} {\bibinfo {author} {\bibfnamefont {C.}~\bibnamefont
  {Adolph}} \emph {et~al.} (\bibinfo {collaboration} {COMPASS}),\ }\href
  {\doibase 10.1016/j.physletb.2017.04.042} {\bibfield  {journal} {\bibinfo
  {journal} {Phys. Lett.}\ }\textbf {\bibinfo {volume} {B770}},\ \bibinfo
  {pages} {138} (\bibinfo {year} {2017})},\ \Eprint
  {http://arxiv.org/abs/1609.07374} {arXiv:1609.07374 [hep-ex]} \BibitemShut
  {NoStop}%
\bibitem [{\citenamefont {Airapetian}\ \emph {et~al.}(2010)\citenamefont
  {Airapetian} \emph {et~al.}}]{Airapetian:2010ds}%
  \BibitemOpen
  \bibfield  {author} {\bibinfo {author} {\bibfnamefont {A.}~\bibnamefont
  {Airapetian}} \emph {et~al.} (\bibinfo {collaboration} {HERMES}),\ }\href
  {\doibase 10.1016/j.physletb.2010.08.012} {\bibfield  {journal} {\bibinfo
  {journal} {Phys.Lett.}\ }\textbf {\bibinfo {volume} {B693}},\ \bibinfo
  {pages} {11} (\bibinfo {year} {2010})},\ \Eprint
  {http://arxiv.org/abs/1006.4221} {arXiv:1006.4221 [hep-ex]} \BibitemShut
  {NoStop}%
\bibitem [{\citenamefont {Adamczyk}\ \emph {et~al.}(2016)\citenamefont
  {Adamczyk} \emph {et~al.}}]{Adamczyk:2015gyk}%
  \BibitemOpen
  \bibfield  {author} {\bibinfo {author} {\bibfnamefont {L.}~\bibnamefont
  {Adamczyk}} \emph {et~al.} (\bibinfo {collaboration} {STAR}),\ }\href
  {\doibase 10.1103/PhysRevLett.116.132301} {\bibfield  {journal} {\bibinfo
  {journal} {Phys. Rev. Lett.}\ }\textbf {\bibinfo {volume} {116}},\ \bibinfo
  {pages} {132301} (\bibinfo {year} {2016})},\ \Eprint
  {http://arxiv.org/abs/1511.06003} {arXiv:1511.06003 [nucl-ex]} \BibitemShut
  {NoStop}%
\bibitem [{\citenamefont {Aghasyan}\ \emph {et~al.}(2017)\citenamefont
  {Aghasyan} \emph {et~al.}}]{Aghasyan:2017jop}%
  \BibitemOpen
  \bibfield  {author} {\bibinfo {author} {\bibfnamefont {M.}~\bibnamefont
  {Aghasyan}} \emph {et~al.} (\bibinfo {collaboration} {COMPASS}),\ }\href
  {\doibase 10.1103/PhysRevLett.119.112002} {\bibfield  {journal} {\bibinfo
  {journal} {Phys. Rev. Lett.}\ }\textbf {\bibinfo {volume} {119}},\ \bibinfo
  {pages} {112002} (\bibinfo {year} {2017})},\ \Eprint
  {http://arxiv.org/abs/1704.00488} {arXiv:1704.00488 [hep-ex]} \BibitemShut
  {NoStop}%
\bibitem [{\citenamefont {Seidl}\ \emph {et~al.}(2008)\citenamefont {Seidl}
  \emph {et~al.}}]{Seidl:2008xc}%
  \BibitemOpen
  \bibfield  {author} {\bibinfo {author} {\bibfnamefont {R.}~\bibnamefont
  {Seidl}} \emph {et~al.} (\bibinfo {collaboration} {Belle}),\ }\href {\doibase
  10.1103/PhysRevD.78.032011} {\bibfield  {journal} {\bibinfo  {journal} {Phys.
  Rev.}\ }\textbf {\bibinfo {volume} {D78}},\ \bibinfo {pages} {032011}
  (\bibinfo {year} {2008})},\ \Eprint {http://arxiv.org/abs/0805.2975}
  {arXiv:0805.2975 [hep-ex]} \BibitemShut {NoStop}%
\bibitem [{\citenamefont {Lees}\ \emph {et~al.}(2014)\citenamefont {Lees} \emph
  {et~al.}}]{TheBABAR:2013yha}%
  \BibitemOpen
  \bibfield  {author} {\bibinfo {author} {\bibfnamefont {J.~P.}\ \bibnamefont
  {Lees}} \emph {et~al.} (\bibinfo {collaboration} {BaBar}),\ }\href {\doibase
  10.1103/PhysRevD.90.052003} {\bibfield  {journal} {\bibinfo  {journal} {Phys.
  Rev.}\ }\textbf {\bibinfo {volume} {D90}},\ \bibinfo {pages} {052003}
  (\bibinfo {year} {2014})},\ \Eprint {http://arxiv.org/abs/1309.5278}
  {arXiv:1309.5278 [hep-ex]} \BibitemShut {NoStop}%
\bibitem [{\citenamefont {Lees}\ \emph {et~al.}(2015)\citenamefont {Lees} \emph
  {et~al.}}]{Aubert:2015hha}%
  \BibitemOpen
  \bibfield  {author} {\bibinfo {author} {\bibfnamefont {J.~P.}\ \bibnamefont
  {Lees}} \emph {et~al.} (\bibinfo {collaboration} {BaBar}),\ }\href {\doibase
  10.1103/PhysRevD.92.111101} {\bibfield  {journal} {\bibinfo  {journal} {Phys.
  Rev.}\ }\textbf {\bibinfo {volume} {D92}},\ \bibinfo {pages} {111101}
  (\bibinfo {year} {2015})},\ \Eprint {http://arxiv.org/abs/1506.05864}
  {arXiv:1506.05864 [hep-ex]} \BibitemShut {NoStop}%
\bibitem [{\citenamefont {Ablikim}\ \emph {et~al.}(2016)\citenamefont {Ablikim}
  \emph {et~al.}}]{Ablikim:2015pta}%
  \BibitemOpen
  \bibfield  {author} {\bibinfo {author} {\bibfnamefont {M.}~\bibnamefont
  {Ablikim}} \emph {et~al.} (\bibinfo {collaboration} {BESIII}),\ }\href
  {\doibase 10.1103/PhysRevLett.116.042001} {\bibfield  {journal} {\bibinfo
  {journal} {Phys. Rev. Lett.}\ }\textbf {\bibinfo {volume} {116}},\ \bibinfo
  {pages} {042001} (\bibinfo {year} {2016})},\ \Eprint
  {http://arxiv.org/abs/1507.06824} {arXiv:1507.06824 [hep-ex]} \BibitemShut
  {NoStop}%
\bibitem [{\citenamefont {Li}\ \emph {et~al.}(2019)\citenamefont {Li} \emph
  {et~al.}}]{Li:2019iyt}%
  \BibitemOpen
  \bibfield  {author} {\bibinfo {author} {\bibfnamefont {H.}~\bibnamefont {Li}}
  \emph {et~al.} (\bibinfo {collaboration} {Belle}),\ }\href {\doibase
  10.1103/PhysRevD.100.092008} {\bibfield  {journal} {\bibinfo  {journal}
  {Phys. Rev.}\ }\textbf {\bibinfo {volume} {D100}},\ \bibinfo {pages} {092008}
  (\bibinfo {year} {2019})},\ \Eprint {http://arxiv.org/abs/1909.01857}
  {arXiv:1909.01857 [hep-ex]} \BibitemShut {NoStop}%
\bibitem [{\citenamefont {Ralston}\ and\ \citenamefont
  {Soper}(1979)}]{Ralston:1979ys}%
  \BibitemOpen
  \bibfield  {author} {\bibinfo {author} {\bibfnamefont {J.~P.}\ \bibnamefont
  {Ralston}}\ and\ \bibinfo {author} {\bibfnamefont {D.~E.}\ \bibnamefont
  {Soper}},\ }\href {\doibase 10.1016/0550-3213(79)90082-8} {\bibfield
  {journal} {\bibinfo  {journal} {Nucl. Phys.}\ }\textbf {\bibinfo {volume}
  {B152}},\ \bibinfo {pages} {109} (\bibinfo {year} {1979})}\BibitemShut
  {NoStop}%
\bibitem [{\citenamefont {Collins}(1993)}]{Collins:1992kk}%
  \BibitemOpen
  \bibfield  {author} {\bibinfo {author} {\bibfnamefont {J.~C.}\ \bibnamefont
  {Collins}},\ }\href {\doibase 10.1016/0550-3213(93)90262-N} {\bibfield
  {journal} {\bibinfo  {journal} {Nucl.~Phys.}\ }\textbf {\bibinfo {volume}
  {B396}},\ \bibinfo {pages} {161} (\bibinfo {year} {1993})},\ \Eprint
  {http://arxiv.org/abs/hep-ph/9208213} {hep-ph/9208213} \BibitemShut {NoStop}%
\bibitem [{\citenamefont {Bacchetta}\ and\ \citenamefont
  {Prokudin}(2013)}]{Bacchetta:2013pqa}%
  \BibitemOpen
  \bibfield  {author} {\bibinfo {author} {\bibfnamefont {A.}~\bibnamefont
  {Bacchetta}}\ and\ \bibinfo {author} {\bibfnamefont {A.}~\bibnamefont
  {Prokudin}},\ }\href {\doibase 10.1016/j.nuclphysb.2013.07.013} {\bibfield
  {journal} {\bibinfo  {journal} {Nucl. Phys.}\ }\textbf {\bibinfo {volume}
  {B875}},\ \bibinfo {pages} {536} (\bibinfo {year} {2013})},\ \Eprint
  {http://arxiv.org/abs/1303.2129} {arXiv:1303.2129 [hep-ph]} \BibitemShut
  {NoStop}%
\bibitem [{\citenamefont {Kang}\ \emph {et~al.}(2016)\citenamefont {Kang},
  \citenamefont {Prokudin}, \citenamefont {Sun},\ and\ \citenamefont
  {Yuan}}]{Kang:2015msa}%
  \BibitemOpen
  \bibfield  {author} {\bibinfo {author} {\bibfnamefont {Z.-B.}\ \bibnamefont
  {Kang}}, \bibinfo {author} {\bibfnamefont {A.}~\bibnamefont {Prokudin}},
  \bibinfo {author} {\bibfnamefont {P.}~\bibnamefont {Sun}}, \ and\ \bibinfo
  {author} {\bibfnamefont {F.}~\bibnamefont {Yuan}},\ }\href {\doibase
  10.1103/PhysRevD.93.014009} {\bibfield  {journal} {\bibinfo  {journal} {Phys.
  Rev.}\ }\textbf {\bibinfo {volume} {D93}},\ \bibinfo {pages} {014009}
  (\bibinfo {year} {2016})},\ \Eprint {http://arxiv.org/abs/1505.05589}
  {arXiv:1505.05589 [hep-ph]} \BibitemShut {NoStop}%
\bibitem [{\citenamefont {Kanazawa}\ \emph {et~al.}(2014)\citenamefont
  {Kanazawa}, \citenamefont {Koike}, \citenamefont {Metz},\ and\ \citenamefont
  {Pitonyak}}]{Kanazawa:2014dca}%
  \BibitemOpen
  \bibfield  {author} {\bibinfo {author} {\bibfnamefont {K.}~\bibnamefont
  {Kanazawa}}, \bibinfo {author} {\bibfnamefont {Y.}~\bibnamefont {Koike}},
  \bibinfo {author} {\bibfnamefont {A.}~\bibnamefont {Metz}}, \ and\ \bibinfo
  {author} {\bibfnamefont {D.}~\bibnamefont {Pitonyak}},\ }\href {\doibase
  10.1103/PhysRevD.89.111501} {\bibfield  {journal} {\bibinfo  {journal} {Phys.
  Rev.}\ }\textbf {\bibinfo {volume} {D89}},\ \bibinfo {pages} {111501(R)}
  (\bibinfo {year} {2014})},\ \Eprint {http://arxiv.org/abs/1404.1033}
  {arXiv:1404.1033 [hep-ph]} \BibitemShut {NoStop}%
\bibitem [{\citenamefont {Gamberg}\ \emph {et~al.}(2017)\citenamefont
  {Gamberg}, \citenamefont {Kang}, \citenamefont {Pitonyak},\ and\
  \citenamefont {Prokudin}}]{Gamberg:2017gle}%
  \BibitemOpen
  \bibfield  {author} {\bibinfo {author} {\bibfnamefont {L.}~\bibnamefont
  {Gamberg}}, \bibinfo {author} {\bibfnamefont {Z.-B.}\ \bibnamefont {Kang}},
  \bibinfo {author} {\bibfnamefont {D.}~\bibnamefont {Pitonyak}}, \ and\
  \bibinfo {author} {\bibfnamefont {A.}~\bibnamefont {Prokudin}},\ }\href
  {\doibase 10.1016/j.physletb.2017.04.061} {\bibfield  {journal} {\bibinfo
  {journal} {Phys. Lett.}\ }\textbf {\bibinfo {volume} {B770}},\ \bibinfo
  {pages} {242} (\bibinfo {year} {2017})},\ \Eprint
  {http://arxiv.org/abs/1701.09170} {arXiv:1701.09170 [hep-ph]} \BibitemShut
  {NoStop}%
\bibitem [{\citenamefont {Anselmino}\ \emph
  {et~al.}(2005{\natexlab{a}})\citenamefont {Anselmino}, \citenamefont
  {Boglione}, \citenamefont {D'Alesio}, \citenamefont {Kotzinian},
  \citenamefont {Murgia} \emph {et~al.}}]{Anselmino:2005nn}%
  \BibitemOpen
  \bibfield  {author} {\bibinfo {author} {\bibfnamefont {M.}~\bibnamefont
  {Anselmino}}, \bibinfo {author} {\bibfnamefont {M.}~\bibnamefont {Boglione}},
  \bibinfo {author} {\bibfnamefont {U.}~\bibnamefont {D'Alesio}}, \bibinfo
  {author} {\bibfnamefont {A.}~\bibnamefont {Kotzinian}}, \bibinfo {author}
  {\bibfnamefont {F.}~\bibnamefont {Murgia}},  \emph {et~al.},\ }\href
  {\doibase 10.1103/PhysRevD.71.074006} {\bibfield  {journal} {\bibinfo
  {journal} {Phys.Rev.}\ }\textbf {\bibinfo {volume} {D71}},\ \bibinfo {pages}
  {074006} (\bibinfo {year} {2005}{\natexlab{a}})},\ \Eprint
  {http://arxiv.org/abs/hep-ph/0501196} {arXiv:hep-ph/0501196 [hep-ph]}
  \BibitemShut {NoStop}%
\bibitem [{\citenamefont {Anselmino}\ \emph {et~al.}(2001)\citenamefont
  {Anselmino}, \citenamefont {Boer}, \citenamefont {D'Alesio},\ and\
  \citenamefont {Murgia}}]{Anselmino:2000vs}%
  \BibitemOpen
  \bibfield  {author} {\bibinfo {author} {\bibfnamefont {M.}~\bibnamefont
  {Anselmino}}, \bibinfo {author} {\bibfnamefont {D.}~\bibnamefont {Boer}},
  \bibinfo {author} {\bibfnamefont {U.}~\bibnamefont {D'Alesio}}, \ and\
  \bibinfo {author} {\bibfnamefont {F.}~\bibnamefont {Murgia}},\ }\href
  {\doibase 10.1103/PhysRevD.63.054029} {\bibfield  {journal} {\bibinfo
  {journal} {Phys. Rev.}\ }\textbf {\bibinfo {volume} {D63}},\ \bibinfo {pages}
  {054029} (\bibinfo {year} {2001})},\ \Eprint
  {http://arxiv.org/abs/hep-ph/0008186} {arXiv:hep-ph/0008186 [hep-ph]}
  \BibitemShut {NoStop}%
\bibitem [{\citenamefont {Anselmino}\ \emph
  {et~al.}(2005{\natexlab{b}})\citenamefont {Anselmino}, \citenamefont
  {Boglione}, \citenamefont {D'Alesio}, \citenamefont {Kotzinian},
  \citenamefont {Murgia} \emph {et~al.}}]{Anselmino:2005ea}%
  \BibitemOpen
  \bibfield  {author} {\bibinfo {author} {\bibfnamefont {M.}~\bibnamefont
  {Anselmino}}, \bibinfo {author} {\bibfnamefont {M.}~\bibnamefont {Boglione}},
  \bibinfo {author} {\bibfnamefont {U.}~\bibnamefont {D'Alesio}}, \bibinfo
  {author} {\bibfnamefont {A.}~\bibnamefont {Kotzinian}}, \bibinfo {author}
  {\bibfnamefont {F.}~\bibnamefont {Murgia}},  \emph {et~al.},\ }\href
  {\doibase 10.1103/PhysRevD.72.094007, 10.1103/PhysRevD.72.099903} {\bibfield
  {journal} {\bibinfo  {journal} {Phys.~Rev.}\ }\textbf {\bibinfo {volume}
  {D72}},\ \bibinfo {pages} {094007} (\bibinfo {year} {2005}{\natexlab{b}})},\
  \Eprint {http://arxiv.org/abs/hep-ph/0507181} {arXiv:hep-ph/0507181 [hep-ph]}
  \BibitemShut {NoStop}%
\bibitem [{\citenamefont {Vogelsang}\ and\ \citenamefont
  {Yuan}(2005)}]{Vogelsang:2005cs}%
  \BibitemOpen
  \bibfield  {author} {\bibinfo {author} {\bibfnamefont {W.}~\bibnamefont
  {Vogelsang}}\ and\ \bibinfo {author} {\bibfnamefont {F.}~\bibnamefont
  {Yuan}},\ }\href@noop {} {\bibfield  {journal} {\bibinfo  {journal} {Phys.
  Rev.}\ }\textbf {\bibinfo {volume} {D72}},\ \bibinfo {pages} {054028}
  (\bibinfo {year} {2005})},\ \Eprint {http://arxiv.org/abs/hep-ph/0507266}
  {hep-ph/0507266} \BibitemShut {NoStop}%
\bibitem [{\citenamefont {Collins}\ \emph {et~al.}(2006)\citenamefont
  {Collins}, \citenamefont {Efremov}, \citenamefont {Goeke}, \citenamefont
  {Menzel}, \citenamefont {Metz} \emph {et~al.}}]{Collins:2005ie}%
  \BibitemOpen
  \bibfield  {author} {\bibinfo {author} {\bibfnamefont {J.}~\bibnamefont
  {Collins}}, \bibinfo {author} {\bibfnamefont {A.}~\bibnamefont {Efremov}},
  \bibinfo {author} {\bibfnamefont {K.}~\bibnamefont {Goeke}}, \bibinfo
  {author} {\bibfnamefont {S.}~\bibnamefont {Menzel}}, \bibinfo {author}
  {\bibfnamefont {A.}~\bibnamefont {Metz}},  \emph {et~al.},\ }\href {\doibase
  10.1103/PhysRevD.73.014021} {\bibfield  {journal} {\bibinfo  {journal}
  {Phys.~Rev.}\ }\textbf {\bibinfo {volume} {D73}},\ \bibinfo {pages} {014021}
  (\bibinfo {year} {2006})},\ \Eprint {http://arxiv.org/abs/hep-ph/0509076}
  {arXiv:hep-ph/0509076 [hep-ph]} \BibitemShut {NoStop}%
\bibitem [{\citenamefont {Collins}\ \emph {et~al.}(2005)\citenamefont {Collins}
  \emph {et~al.}}]{Collins:2005rq}%
  \BibitemOpen
  \bibfield  {author} {\bibinfo {author} {\bibfnamefont {J.~C.}\ \bibnamefont
  {Collins}} \emph {et~al.},\ }\href@noop {} {\  (\bibinfo {year} {2005})},\
  \Eprint {http://arxiv.org/abs/hep-ph/0511272} {hep-ph/0511272} \BibitemShut
  {NoStop}%
\bibitem [{\citenamefont {Anselmino}\ \emph {et~al.}(2007)\citenamefont
  {Anselmino}, \citenamefont {Boglione}, \citenamefont {D'Alesio},
  \citenamefont {Kotzinian}, \citenamefont {Murgia} \emph
  {et~al.}}]{Anselmino:2007fs}%
  \BibitemOpen
  \bibfield  {author} {\bibinfo {author} {\bibfnamefont {M.}~\bibnamefont
  {Anselmino}}, \bibinfo {author} {\bibfnamefont {M.}~\bibnamefont {Boglione}},
  \bibinfo {author} {\bibfnamefont {U.}~\bibnamefont {D'Alesio}}, \bibinfo
  {author} {\bibfnamefont {A.}~\bibnamefont {Kotzinian}}, \bibinfo {author}
  {\bibfnamefont {F.}~\bibnamefont {Murgia}},  \emph {et~al.},\ }\href
  {\doibase 10.1103/PhysRevD.75.054032} {\bibfield  {journal} {\bibinfo
  {journal} {Phys.~Rev.}\ }\textbf {\bibinfo {volume} {D75}},\ \bibinfo {pages}
  {054032} (\bibinfo {year} {2007})},\ \Eprint
  {http://arxiv.org/abs/hep-ph/0701006} {arXiv:hep-ph/0701006 [hep-ph]}
  \BibitemShut {NoStop}%
\bibitem [{\citenamefont {Anselmino}\ \emph {et~al.}(2009)\citenamefont
  {Anselmino}, \citenamefont {Boglione}, \citenamefont {D'Alesio},
  \citenamefont {Kotzinian}, \citenamefont {Murgia}, \citenamefont {Prokudin},\
  and\ \citenamefont {Melis}}]{Anselmino:2008jk}%
  \BibitemOpen
  \bibfield  {author} {\bibinfo {author} {\bibfnamefont {M.}~\bibnamefont
  {Anselmino}}, \bibinfo {author} {\bibfnamefont {M.}~\bibnamefont {Boglione}},
  \bibinfo {author} {\bibfnamefont {U.}~\bibnamefont {D'Alesio}}, \bibinfo
  {author} {\bibfnamefont {A.}~\bibnamefont {Kotzinian}}, \bibinfo {author}
  {\bibfnamefont {F.}~\bibnamefont {Murgia}}, \bibinfo {author} {\bibfnamefont
  {A.}~\bibnamefont {Prokudin}}, \ and\ \bibinfo {author} {\bibfnamefont
  {S.}~\bibnamefont {Melis}},\ }\bibfield  {booktitle} {\emph {\bibinfo
  {booktitle} {{Proceedings, Ringberg Workshop on New Trends in HERA Physics
  2008: Ringberg Castle, Tegernsee, Germany, 5-10 October 2008}}},\ }\href
  {\doibase 10.1016/j.nuclphysbps.2009.03.117} {\bibfield  {journal} {\bibinfo
  {journal} {Nucl. Phys. Proc. Suppl.}\ }\textbf {\bibinfo {volume} {191}},\
  \bibinfo {pages} {98} (\bibinfo {year} {2009})},\ \Eprint
  {http://arxiv.org/abs/0812.4366} {arXiv:0812.4366 [hep-ph]} \BibitemShut
  {NoStop}%
\bibitem [{\citenamefont {Schweitzer}\ \emph {et~al.}(2010)\citenamefont
  {Schweitzer}, \citenamefont {Teckentrup},\ and\ \citenamefont
  {Metz}}]{Schweitzer:2010tt}%
  \BibitemOpen
  \bibfield  {author} {\bibinfo {author} {\bibfnamefont {P.}~\bibnamefont
  {Schweitzer}}, \bibinfo {author} {\bibfnamefont {T.}~\bibnamefont
  {Teckentrup}}, \ and\ \bibinfo {author} {\bibfnamefont {A.}~\bibnamefont
  {Metz}},\ }\href {\doibase 10.1103/PhysRevD.81.094019} {\bibfield  {journal}
  {\bibinfo  {journal} {Phys. Rev. D}\ }\textbf {\bibinfo {volume} {81}},\
  \bibinfo {pages} {094019} (\bibinfo {year} {2010})},\ \Eprint
  {http://arxiv.org/abs/1003.2190} {arXiv:1003.2190 [hep-ph]} \BibitemShut
  {NoStop}%
\bibitem [{\citenamefont {Qiu}\ \emph {et~al.}(2011)\citenamefont {Qiu},
  \citenamefont {Schlegel},\ and\ \citenamefont {Vogelsang}}]{Qiu:2011ai}%
  \BibitemOpen
  \bibfield  {author} {\bibinfo {author} {\bibfnamefont {J.-W.}\ \bibnamefont
  {Qiu}}, \bibinfo {author} {\bibfnamefont {M.}~\bibnamefont {Schlegel}}, \
  and\ \bibinfo {author} {\bibfnamefont {W.}~\bibnamefont {Vogelsang}},\ }\href
  {\doibase 10.1103/PhysRevLett.107.062001} {\bibfield  {journal} {\bibinfo
  {journal} {Phys. Rev. Lett.}\ }\textbf {\bibinfo {volume} {107}},\ \bibinfo
  {pages} {062001} (\bibinfo {year} {2011})},\ \Eprint
  {http://arxiv.org/abs/1103.3861} {arXiv:1103.3861 [hep-ph]} \BibitemShut
  {NoStop}%
\bibitem [{\citenamefont {Anselmino}\ \emph {et~al.}(2013)\citenamefont
  {Anselmino}, \citenamefont {Boglione}, \citenamefont {D'Alesio},
  \citenamefont {Melis}, \citenamefont {Murgia} \emph
  {et~al.}}]{Anselmino:2013vqa}%
  \BibitemOpen
  \bibfield  {author} {\bibinfo {author} {\bibfnamefont {M.}~\bibnamefont
  {Anselmino}}, \bibinfo {author} {\bibfnamefont {M.}~\bibnamefont {Boglione}},
  \bibinfo {author} {\bibfnamefont {U.}~\bibnamefont {D'Alesio}}, \bibinfo
  {author} {\bibfnamefont {S.}~\bibnamefont {Melis}}, \bibinfo {author}
  {\bibfnamefont {F.}~\bibnamefont {Murgia}},  \emph {et~al.},\ }\href
  {\doibase 10.1103/PhysRevD.87.094019} {\bibfield  {journal} {\bibinfo
  {journal} {Phys.~Rev.}\ }\textbf {\bibinfo {volume} {D87}},\ \bibinfo {pages}
  {094019} (\bibinfo {year} {2013})},\ \Eprint {http://arxiv.org/abs/1303.3822}
  {arXiv:1303.3822 [hep-ph]} \BibitemShut {NoStop}%
\bibitem [{\citenamefont {Signori}\ \emph {et~al.}(2013)\citenamefont
  {Signori}, \citenamefont {Bacchetta}, \citenamefont {Radici},\ and\
  \citenamefont {Schnell}}]{Signori:2013mda}%
  \BibitemOpen
  \bibfield  {author} {\bibinfo {author} {\bibfnamefont {A.}~\bibnamefont
  {Signori}}, \bibinfo {author} {\bibfnamefont {A.}~\bibnamefont {Bacchetta}},
  \bibinfo {author} {\bibfnamefont {M.}~\bibnamefont {Radici}}, \ and\ \bibinfo
  {author} {\bibfnamefont {G.}~\bibnamefont {Schnell}},\ }\href {\doibase
  10.1007/JHEP11(2013)194} {\bibfield  {journal} {\bibinfo  {journal} {JHEP}\
  }\textbf {\bibinfo {volume} {11}},\ \bibinfo {pages} {194} (\bibinfo {year}
  {2013})},\ \Eprint {http://arxiv.org/abs/1309.3507} {arXiv:1309.3507
  [hep-ph]} \BibitemShut {NoStop}%
\bibitem [{\citenamefont {Anselmino}\ \emph {et~al.}(2014)\citenamefont
  {Anselmino}, \citenamefont {Boglione}, \citenamefont {Gonzalez~Hernandez},
  \citenamefont {Melis},\ and\ \citenamefont {Prokudin}}]{Anselmino:2013lza}%
  \BibitemOpen
  \bibfield  {author} {\bibinfo {author} {\bibfnamefont {M.}~\bibnamefont
  {Anselmino}}, \bibinfo {author} {\bibfnamefont {M.}~\bibnamefont {Boglione}},
  \bibinfo {author} {\bibfnamefont {J.~O.}\ \bibnamefont {Gonzalez~Hernandez}},
  \bibinfo {author} {\bibfnamefont {S.}~\bibnamefont {Melis}}, \ and\ \bibinfo
  {author} {\bibfnamefont {A.}~\bibnamefont {Prokudin}},\ }\href {\doibase
  10.1007/JHEP04(2014)005} {\bibfield  {journal} {\bibinfo  {journal} {JHEP}\
  }\textbf {\bibinfo {volume} {04}},\ \bibinfo {pages} {005} (\bibinfo {year}
  {2014})},\ \Eprint {http://arxiv.org/abs/1312.6261} {arXiv:1312.6261
  [hep-ph]} \BibitemShut {NoStop}%
\bibitem [{\citenamefont {Boer}\ and\ \citenamefont {den
  Dunnen}(2014)}]{Boer:2014tka}%
  \BibitemOpen
  \bibfield  {author} {\bibinfo {author} {\bibfnamefont {D.}~\bibnamefont
  {Boer}}\ and\ \bibinfo {author} {\bibfnamefont {W.~J.}\ \bibnamefont {den
  Dunnen}},\ }\href {\doibase 10.1016/j.nuclphysb.2014.07.006} {\bibfield
  {journal} {\bibinfo  {journal} {Nucl. Phys. B}\ }\textbf {\bibinfo {volume}
  {886}},\ \bibinfo {pages} {421} (\bibinfo {year} {2014})},\ \Eprint
  {http://arxiv.org/abs/1404.6753} {arXiv:1404.6753 [hep-ph]} \BibitemShut
  {NoStop}%
\bibitem [{\citenamefont {D'Alesio}\ \emph
  {et~al.}(2020{\natexlab{a}})\citenamefont {D'Alesio}, \citenamefont
  {Murgia},\ and\ \citenamefont {Zaccheddu}}]{DAlesio:2020wjq}%
  \BibitemOpen
  \bibfield  {author} {\bibinfo {author} {\bibfnamefont {U.}~\bibnamefont
  {D'Alesio}}, \bibinfo {author} {\bibfnamefont {F.}~\bibnamefont {Murgia}}, \
  and\ \bibinfo {author} {\bibfnamefont {M.}~\bibnamefont {Zaccheddu}},\
  }\href@noop {} {\  (\bibinfo {year} {2020}{\natexlab{a}})},\ \Eprint
  {http://arxiv.org/abs/2003.01128} {arXiv:2003.01128 [hep-ph]} \BibitemShut
  {NoStop}%
\bibitem [{\citenamefont {Callos}\ \emph {et~al.}(2020)\citenamefont {Callos},
  \citenamefont {Kang},\ and\ \citenamefont {Terry}}]{Callos:2020qtu}%
  \BibitemOpen
  \bibfield  {author} {\bibinfo {author} {\bibfnamefont {D.}~\bibnamefont
  {Callos}}, \bibinfo {author} {\bibfnamefont {Z.-B.}\ \bibnamefont {Kang}}, \
  and\ \bibinfo {author} {\bibfnamefont {J.}~\bibnamefont {Terry}},\
  }\href@noop {} {\  (\bibinfo {year} {2020})},\ \Eprint
  {http://arxiv.org/abs/2003.04828} {arXiv:2003.04828 [hep-ph]} \BibitemShut
  {NoStop}%
\bibitem [{\citenamefont {Orginos}\ \emph {et~al.}(2017)\citenamefont
  {Orginos}, \citenamefont {Radyushkin}, \citenamefont {Karpie},\ and\
  \citenamefont {Zafeiropoulos}}]{Orginos:2017kos}%
  \BibitemOpen
  \bibfield  {author} {\bibinfo {author} {\bibfnamefont {K.}~\bibnamefont
  {Orginos}}, \bibinfo {author} {\bibfnamefont {A.}~\bibnamefont {Radyushkin}},
  \bibinfo {author} {\bibfnamefont {J.}~\bibnamefont {Karpie}}, \ and\ \bibinfo
  {author} {\bibfnamefont {S.}~\bibnamefont {Zafeiropoulos}},\ }\href {\doibase
  10.1103/PhysRevD.96.094503} {\bibfield  {journal} {\bibinfo  {journal} {Phys.
  Rev. D}\ }\textbf {\bibinfo {volume} {96}},\ \bibinfo {pages} {094503}
  (\bibinfo {year} {2017})},\ \Eprint {http://arxiv.org/abs/1706.05373}
  {arXiv:1706.05373 [hep-ph]} \BibitemShut {NoStop}%
\bibitem [{\citenamefont {Anselmino}\ \emph {et~al.}(2017)\citenamefont
  {Anselmino}, \citenamefont {Boglione}, \citenamefont {D'Alesio},
  \citenamefont {Murgia},\ and\ \citenamefont {Prokudin}}]{Anselmino:2016uie}%
  \BibitemOpen
  \bibfield  {author} {\bibinfo {author} {\bibfnamefont {M.}~\bibnamefont
  {Anselmino}}, \bibinfo {author} {\bibfnamefont {M.}~\bibnamefont {Boglione}},
  \bibinfo {author} {\bibfnamefont {U.}~\bibnamefont {D'Alesio}}, \bibinfo
  {author} {\bibfnamefont {F.}~\bibnamefont {Murgia}}, \ and\ \bibinfo {author}
  {\bibfnamefont {A.}~\bibnamefont {Prokudin}},\ }\href {\doibase
  10.1007/JHEP04(2017)046} {\bibfield  {journal} {\bibinfo  {journal} {JHEP}\
  }\textbf {\bibinfo {volume} {04}},\ \bibinfo {pages} {046} (\bibinfo {year}
  {2017})},\ \Eprint {http://arxiv.org/abs/1612.06413} {arXiv:1612.06413
  [hep-ph]} \BibitemShut {NoStop}%
\bibitem [{\citenamefont {Anselmino}\ \emph {et~al.}(2015)\citenamefont
  {Anselmino}, \citenamefont {Boglione}, \citenamefont {D'Alesio},
  \citenamefont {Gonzalez~Hernandez}, \citenamefont {Melis}, \citenamefont
  {Murgia},\ and\ \citenamefont {Prokudin}}]{Anselmino:2015sxa}%
  \BibitemOpen
  \bibfield  {author} {\bibinfo {author} {\bibfnamefont {M.}~\bibnamefont
  {Anselmino}}, \bibinfo {author} {\bibfnamefont {M.}~\bibnamefont {Boglione}},
  \bibinfo {author} {\bibfnamefont {U.}~\bibnamefont {D'Alesio}}, \bibinfo
  {author} {\bibfnamefont {J.~O.}\ \bibnamefont {Gonzalez~Hernandez}}, \bibinfo
  {author} {\bibfnamefont {S.}~\bibnamefont {Melis}}, \bibinfo {author}
  {\bibfnamefont {F.}~\bibnamefont {Murgia}}, \ and\ \bibinfo {author}
  {\bibfnamefont {A.}~\bibnamefont {Prokudin}},\ }\href {\doibase
  10.1103/PhysRevD.92.114023} {\bibfield  {journal} {\bibinfo  {journal} {Phys.
  Rev.}\ }\textbf {\bibinfo {volume} {D92}},\ \bibinfo {pages} {114023}
  (\bibinfo {year} {2015})},\ \Eprint {http://arxiv.org/abs/1510.05389}
  {arXiv:1510.05389 [hep-ph]} \BibitemShut {NoStop}%
\bibitem [{\citenamefont {Sun}\ and\ \citenamefont {Yuan}(2013)}]{Sun:2013dya}%
  \BibitemOpen
  \bibfield  {author} {\bibinfo {author} {\bibfnamefont {P.}~\bibnamefont
  {Sun}}\ and\ \bibinfo {author} {\bibfnamefont {F.}~\bibnamefont {Yuan}},\
  }\href {\doibase 10.1103/PhysRevD.88.034016} {\bibfield  {journal} {\bibinfo
  {journal} {Phys. Rev.}\ }\textbf {\bibinfo {volume} {D88}},\ \bibinfo {pages}
  {034016} (\bibinfo {year} {2013})},\ \Eprint {http://arxiv.org/abs/1304.5037}
  {arXiv:1304.5037 [hep-ph]} \BibitemShut {NoStop}%
\bibitem [{\citenamefont {Kang}\ \emph {et~al.}(2015)\citenamefont {Kang},
  \citenamefont {Prokudin}, \citenamefont {Sun},\ and\ \citenamefont
  {Yuan}}]{Kang:2014zza}%
  \BibitemOpen
  \bibfield  {author} {\bibinfo {author} {\bibfnamefont {Z.-B.}\ \bibnamefont
  {Kang}}, \bibinfo {author} {\bibfnamefont {A.}~\bibnamefont {Prokudin}},
  \bibinfo {author} {\bibfnamefont {P.}~\bibnamefont {Sun}}, \ and\ \bibinfo
  {author} {\bibfnamefont {F.}~\bibnamefont {Yuan}},\ }\href {\doibase
  10.1103/PhysRevD.91.071501} {\bibfield  {journal} {\bibinfo  {journal} {Phys.
  Rev.}\ }\textbf {\bibinfo {volume} {D91}},\ \bibinfo {pages} {071501}
  (\bibinfo {year} {2015})},\ \Eprint {http://arxiv.org/abs/1410.4877}
  {arXiv:1410.4877 [hep-ph]} \BibitemShut {NoStop}%
\bibitem [{\citenamefont {Echevarria}\ \emph {et~al.}(2014)\citenamefont
  {Echevarria}, \citenamefont {Idilbi}, \citenamefont {Kang},\ and\
  \citenamefont {Vitev}}]{Echevarria:2014xaa}%
  \BibitemOpen
  \bibfield  {author} {\bibinfo {author} {\bibfnamefont {M.~G.}\ \bibnamefont
  {Echevarria}}, \bibinfo {author} {\bibfnamefont {A.}~\bibnamefont {Idilbi}},
  \bibinfo {author} {\bibfnamefont {Z.-B.}\ \bibnamefont {Kang}}, \ and\
  \bibinfo {author} {\bibfnamefont {I.}~\bibnamefont {Vitev}},\ }\href
  {\doibase 10.1103/PhysRevD.89.074013} {\bibfield  {journal} {\bibinfo
  {journal} {Phys.~Rev.}\ }\textbf {\bibinfo {volume} {D89}},\ \bibinfo {pages}
  {074013} (\bibinfo {year} {2014})},\ \Eprint {http://arxiv.org/abs/1401.5078}
  {arXiv:1401.5078 [hep-ph]} \BibitemShut {NoStop}%
\bibitem [{\citenamefont {Kang}\ \emph {et~al.}(2017)\citenamefont {Kang},
  \citenamefont {Prokudin}, \citenamefont {Ringer},\ and\ \citenamefont
  {Yuan}}]{Kang:2017btw}%
  \BibitemOpen
  \bibfield  {author} {\bibinfo {author} {\bibfnamefont {Z.-B.}\ \bibnamefont
  {Kang}}, \bibinfo {author} {\bibfnamefont {A.}~\bibnamefont {Prokudin}},
  \bibinfo {author} {\bibfnamefont {F.}~\bibnamefont {Ringer}}, \ and\ \bibinfo
  {author} {\bibfnamefont {F.}~\bibnamefont {Yuan}},\ }\href {\doibase
  10.1016/j.physletb.2017.10.031} {\bibfield  {journal} {\bibinfo  {journal}
  {Phys. Lett.}\ }\textbf {\bibinfo {volume} {B774}},\ \bibinfo {pages} {635}
  (\bibinfo {year} {2017})},\ \Eprint {http://arxiv.org/abs/1707.00913}
  {arXiv:1707.00913 [hep-ph]} \BibitemShut {NoStop}%
\bibitem [{\citenamefont {Duke}\ and\ \citenamefont
  {Owens}(1984)}]{Duke:1983gd}%
  \BibitemOpen
  \bibfield  {author} {\bibinfo {author} {\bibfnamefont {D.~W.}\ \bibnamefont
  {Duke}}\ and\ \bibinfo {author} {\bibfnamefont {J.~F.}\ \bibnamefont
  {Owens}},\ }\href {\doibase 10.1103/PhysRevD.30.49} {\bibfield  {journal}
  {\bibinfo  {journal} {Phys. Rev.}\ }\textbf {\bibinfo {volume} {D30}},\
  \bibinfo {pages} {49} (\bibinfo {year} {1984})}\BibitemShut {NoStop}%
\bibitem [{\citenamefont {Accardi}\ \emph
  {et~al.}(2016{\natexlab{b}})\citenamefont {Accardi}, \citenamefont {Brady},
  \citenamefont {Melnitchouk}, \citenamefont {Owens},\ and\ \citenamefont
  {Sato}}]{Accardi:2016qay}%
  \BibitemOpen
  \bibfield  {author} {\bibinfo {author} {\bibfnamefont {A.}~\bibnamefont
  {Accardi}}, \bibinfo {author} {\bibfnamefont {L.~T.}\ \bibnamefont {Brady}},
  \bibinfo {author} {\bibfnamefont {W.}~\bibnamefont {Melnitchouk}}, \bibinfo
  {author} {\bibfnamefont {J.~F.}\ \bibnamefont {Owens}}, \ and\ \bibinfo
  {author} {\bibfnamefont {N.}~\bibnamefont {Sato}},\ }\href {\doibase
  10.1103/PhysRevD.93.114017} {\bibfield  {journal} {\bibinfo  {journal} {Phys.
  Rev.}\ }\textbf {\bibinfo {volume} {D93}},\ \bibinfo {pages} {114017}
  (\bibinfo {year} {2016}{\natexlab{b}})},\ \Eprint
  {http://arxiv.org/abs/1602.03154} {arXiv:1602.03154 [hep-ph]} \BibitemShut
  {NoStop}%
\bibitem [{\citenamefont {de~Florian}\ \emph {et~al.}(2007)\citenamefont
  {de~Florian}, \citenamefont {Sassot},\ and\ \citenamefont
  {Stratmann}}]{deFlorian:2007ekg}%
  \BibitemOpen
  \bibfield  {author} {\bibinfo {author} {\bibfnamefont {D.}~\bibnamefont
  {de~Florian}}, \bibinfo {author} {\bibfnamefont {R.}~\bibnamefont {Sassot}},
  \ and\ \bibinfo {author} {\bibfnamefont {M.}~\bibnamefont {Stratmann}},\
  }\href {\doibase 10.1103/PhysRevD.76.074033} {\bibfield  {journal} {\bibinfo
  {journal} {Phys. Rev.}\ }\textbf {\bibinfo {volume} {D76}},\ \bibinfo {pages}
  {074033} (\bibinfo {year} {2007})},\ \Eprint {http://arxiv.org/abs/0707.1506}
  {arXiv:0707.1506 [hep-ph]} \BibitemShut {NoStop}%
\bibitem [{\citenamefont {Barry}\ \emph {et~al.}(2018)\citenamefont {Barry},
  \citenamefont {Sato}, \citenamefont {Melnitchouk},\ and\ \citenamefont
  {Ji}}]{Barry:2018ort}%
  \BibitemOpen
  \bibfield  {author} {\bibinfo {author} {\bibfnamefont {P.~C.}\ \bibnamefont
  {Barry}}, \bibinfo {author} {\bibfnamefont {N.}~\bibnamefont {Sato}},
  \bibinfo {author} {\bibfnamefont {W.}~\bibnamefont {Melnitchouk}}, \ and\
  \bibinfo {author} {\bibfnamefont {C.-R.}\ \bibnamefont {Ji}},\ }\href
  {\doibase 10.1103/PhysRevLett.121.152001} {\bibfield  {journal} {\bibinfo
  {journal} {Phys. Rev. Lett.}\ }\textbf {\bibinfo {volume} {121}},\ \bibinfo
  {pages} {152001} (\bibinfo {year} {2018})},\ \Eprint
  {http://arxiv.org/abs/1804.01965} {arXiv:1804.01965 [hep-ph]} \BibitemShut
  {NoStop}%
\bibitem [{Note1()}]{Note1}%
  \BibitemOpen
  \bibinfo {note} {The precision of the COMPASS Drell-Yan data is such that
  using next-to-leading order pion PDFs will not affect our
  results.}\BibitemShut {Stop}%
\bibitem [{\citenamefont {Schnell}(2010)}]{Schnell:2010zza}%
  \BibitemOpen
  \bibfield  {author} {\bibinfo {author} {\bibfnamefont {G.}~\bibnamefont
  {Schnell}} (\bibinfo {collaboration} {HERMES}),\ }\bibfield  {booktitle}
  {\emph {\bibinfo {booktitle} {{Proceedings, 18th International Workshop on
  Deep-inelastic scattering and related subjects (DIS 2010): Florence, Italy,
  April 19-23, 2010}}},\ }\href {\doibase 10.22323/1.106.0247} {\bibfield
  {journal} {\bibinfo  {journal} {PoS}\ }\textbf {\bibinfo {volume}
  {DIS2010}},\ \bibinfo {pages} {247} (\bibinfo {year} {2010})}\BibitemShut
  {NoStop}%
\bibitem [{\citenamefont {Parsamyan}(2013)}]{Parsamyan:2013fia}%
  \BibitemOpen
  \bibfield  {author} {\bibinfo {author} {\bibfnamefont {B.}~\bibnamefont
  {Parsamyan}},\ }\bibfield  {booktitle} {\emph {\bibinfo {booktitle}
  {{Proceedings, 21st International Workshop on Deep-Inelastic Scattering and
  Related Subjects (DIS 2013): Marseilles, France, April 22-26, 2013}}},\
  }\href {\doibase 10.22323/1.191.0231} {\bibfield  {journal} {\bibinfo
  {journal} {PoS}\ }\textbf {\bibinfo {volume} {DIS2013}},\ \bibinfo {pages}
  {231} (\bibinfo {year} {2013})},\ \Eprint {http://arxiv.org/abs/1307.0183}
  {arXiv:1307.0183 [hep-ex]} \BibitemShut {NoStop}%
\bibitem [{\citenamefont {D'Alesio}\ \emph
  {et~al.}(2020{\natexlab{b}})\citenamefont {D'Alesio}, \citenamefont {Flore},\
  and\ \citenamefont {Prokudin}}]{DAlesio:2020vtw}%
  \BibitemOpen
  \bibfield  {author} {\bibinfo {author} {\bibfnamefont {U.}~\bibnamefont
  {D'Alesio}}, \bibinfo {author} {\bibfnamefont {C.}~\bibnamefont {Flore}}, \
  and\ \bibinfo {author} {\bibfnamefont {A.}~\bibnamefont {Prokudin}},\
  }\href@noop {} {\  (\bibinfo {year} {2020}{\natexlab{b}})},\ \Eprint
  {http://arxiv.org/abs/2001.01573} {arXiv:2001.01573 [hep-ph]} \BibitemShut
  {NoStop}%
\bibitem [{\citenamefont {Airapetian}\ \emph {et~al.}(2013)\citenamefont
  {Airapetian} \emph {et~al.}}]{Airapetian:2012ki}%
  \BibitemOpen
  \bibfield  {author} {\bibinfo {author} {\bibfnamefont {A.}~\bibnamefont
  {Airapetian}} \emph {et~al.} (\bibinfo {collaboration} {HERMES}),\ }\href
  {\doibase 10.1103/PhysRevD.87.074029} {\bibfield  {journal} {\bibinfo
  {journal} {Phys. Rev.}\ }\textbf {\bibinfo {volume} {D87}},\ \bibinfo {pages}
  {074029} (\bibinfo {year} {2013})},\ \Eprint {http://arxiv.org/abs/1212.5407}
  {arXiv:1212.5407 [hep-ex]} \BibitemShut {NoStop}%
\bibitem [{\citenamefont {Sato}\ \emph {et~al.}(2019)\citenamefont {Sato},
  \citenamefont {Andres}, \citenamefont {Ethier},\ and\ \citenamefont
  {Melnitchouk}}]{Sato:2019yez}%
  \BibitemOpen
  \bibfield  {author} {\bibinfo {author} {\bibfnamefont {N.}~\bibnamefont
  {Sato}}, \bibinfo {author} {\bibfnamefont {C.}~\bibnamefont {Andres}},
  \bibinfo {author} {\bibfnamefont {J.~J.}\ \bibnamefont {Ethier}}, \ and\
  \bibinfo {author} {\bibfnamefont {W.}~\bibnamefont {Melnitchouk}} (\bibinfo
  {collaboration} {JAM}),\ }\href@noop {} {\  (\bibinfo {year} {2019})},\
  \Eprint {http://arxiv.org/abs/1905.03788} {arXiv:1905.03788 [hep-ph]}
  \BibitemShut {NoStop}%
\bibitem [{\citenamefont {Radici}\ and\ \citenamefont
  {Bacchetta}(2018)}]{Radici:2018iag}%
  \BibitemOpen
  \bibfield  {author} {\bibinfo {author} {\bibfnamefont {M.}~\bibnamefont
  {Radici}}\ and\ \bibinfo {author} {\bibfnamefont {A.}~\bibnamefont
  {Bacchetta}},\ }\href {\doibase 10.1103/PhysRevLett.120.192001} {\bibfield
  {journal} {\bibinfo  {journal} {Phys. Rev. Lett.}\ }\textbf {\bibinfo
  {volume} {120}},\ \bibinfo {pages} {192001} (\bibinfo {year} {2018})},\
  \Eprint {http://arxiv.org/abs/1802.05212} {arXiv:1802.05212 [hep-ph]}
  \BibitemShut {NoStop}%
\bibitem [{\citenamefont {Benel}\ \emph {et~al.}(2019)\citenamefont {Benel},
  \citenamefont {Courtoy},\ and\ \citenamefont
  {Ferro-Hernandez}}]{Benel:2019mcq}%
  \BibitemOpen
  \bibfield  {author} {\bibinfo {author} {\bibfnamefont {J.}~\bibnamefont
  {Benel}}, \bibinfo {author} {\bibfnamefont {A.}~\bibnamefont {Courtoy}}, \
  and\ \bibinfo {author} {\bibfnamefont {R.}~\bibnamefont {Ferro-Hernandez}},\
  }\href@noop {} {\  (\bibinfo {year} {2019})},\ \Eprint
  {http://arxiv.org/abs/1912.03289} {arXiv:1912.03289 [hep-ph]} \BibitemShut
  {NoStop}%
\bibitem [{JAM()}]{JAM20:code}%
  \BibitemOpen
  \href@noop {} {\bibinfo  {journal}
  {\url{https://github.com/QCDHUB/jam3dlib}}\ }\BibitemShut {NoStop}%
\bibitem [{\citenamefont {Gupta}\ \emph {et~al.}(2018)\citenamefont {Gupta},
  \citenamefont {Jang}, \citenamefont {Yoon}, \citenamefont {Lin},
  \citenamefont {Cirigliano},\ and\ \citenamefont
  {Bhattacharya}}]{Gupta:2018qil}%
  \BibitemOpen
\bibfield  {journal} {  }\bibfield  {author} {\bibinfo {author} {\bibfnamefont
  {R.}~\bibnamefont {Gupta}}, \bibinfo {author} {\bibfnamefont {Y.-C.}\
  \bibnamefont {Jang}}, \bibinfo {author} {\bibfnamefont {B.}~\bibnamefont
  {Yoon}}, \bibinfo {author} {\bibfnamefont {H.-W.}\ \bibnamefont {Lin}},
  \bibinfo {author} {\bibfnamefont {V.}~\bibnamefont {Cirigliano}}, \ and\
  \bibinfo {author} {\bibfnamefont {T.}~\bibnamefont {Bhattacharya}},\ }\href
  {\doibase 10.1103/PhysRevD.98.034503} {\bibfield  {journal} {\bibinfo
  {journal} {Phys. Rev.}\ }\textbf {\bibinfo {volume} {D98}},\ \bibinfo {pages}
  {034503} (\bibinfo {year} {2018})},\ \Eprint
  {http://arxiv.org/abs/1806.09006} {arXiv:1806.09006 [hep-lat]} \BibitemShut
  {NoStop}%
\bibitem [{\citenamefont {Hasan}\ \emph {et~al.}(2019)\citenamefont {Hasan},
  \citenamefont {Green}, \citenamefont {Meinel}, \citenamefont {Engelhardt},
  \citenamefont {Krieg}, \citenamefont {Negele}, \citenamefont {Pochinsky},\
  and\ \citenamefont {Syritsyn}}]{Hasan:2019noy}%
  \BibitemOpen
  \bibfield  {author} {\bibinfo {author} {\bibfnamefont {N.}~\bibnamefont
  {Hasan}}, \bibinfo {author} {\bibfnamefont {J.}~\bibnamefont {Green}},
  \bibinfo {author} {\bibfnamefont {S.}~\bibnamefont {Meinel}}, \bibinfo
  {author} {\bibfnamefont {M.}~\bibnamefont {Engelhardt}}, \bibinfo {author}
  {\bibfnamefont {S.}~\bibnamefont {Krieg}}, \bibinfo {author} {\bibfnamefont
  {J.}~\bibnamefont {Negele}}, \bibinfo {author} {\bibfnamefont
  {A.}~\bibnamefont {Pochinsky}}, \ and\ \bibinfo {author} {\bibfnamefont
  {S.}~\bibnamefont {Syritsyn}},\ }\href {\doibase 10.1103/PhysRevD.99.114505}
  {\bibfield  {journal} {\bibinfo  {journal} {Phys. Rev.}\ }\textbf {\bibinfo
  {volume} {D99}},\ \bibinfo {pages} {114505} (\bibinfo {year} {2019})},\
  \Eprint {http://arxiv.org/abs/1903.06487} {arXiv:1903.06487 [hep-lat]}
  \BibitemShut {NoStop}%
\bibitem [{\citenamefont {Alexandrou}\ \emph {et~al.}(2019)\citenamefont
  {Alexandrou}, \citenamefont {Bacchio}, \citenamefont {Constantinou},
  \citenamefont {Finkenrath}, \citenamefont {Hadjiyiannakou}, \citenamefont
  {Jansen}, \citenamefont {Koutsou},\ and\ \citenamefont {Vaquero
  Aviles-Casco}}]{Alexandrou:2019brg}%
  \BibitemOpen
  \bibfield  {author} {\bibinfo {author} {\bibfnamefont {C.}~\bibnamefont
  {Alexandrou}}, \bibinfo {author} {\bibfnamefont {S.}~\bibnamefont {Bacchio}},
  \bibinfo {author} {\bibfnamefont {M.}~\bibnamefont {Constantinou}}, \bibinfo
  {author} {\bibfnamefont {J.}~\bibnamefont {Finkenrath}}, \bibinfo {author}
  {\bibfnamefont {K.}~\bibnamefont {Hadjiyiannakou}}, \bibinfo {author}
  {\bibfnamefont {K.}~\bibnamefont {Jansen}}, \bibinfo {author} {\bibfnamefont
  {G.}~\bibnamefont {Koutsou}}, \ and\ \bibinfo {author} {\bibfnamefont
  {A.}~\bibnamefont {Vaquero Aviles-Casco}},\ }\href@noop {} {\  (\bibinfo
  {year} {2019})},\ \Eprint {http://arxiv.org/abs/1909.00485} {arXiv:1909.00485
  [hep-lat]} \BibitemShut {NoStop}%
\bibitem [{\citenamefont {Goldstein}\ \emph {et~al.}(2014)\citenamefont
  {Goldstein}, \citenamefont {Gonzalez~Hernandez},\ and\ \citenamefont
  {Liuti}}]{Goldstein:2014aja}%
  \BibitemOpen
  \bibfield  {author} {\bibinfo {author} {\bibfnamefont {G.~R.}\ \bibnamefont
  {Goldstein}}, \bibinfo {author} {\bibfnamefont {J.~O.}\ \bibnamefont
  {Gonzalez~Hernandez}}, \ and\ \bibinfo {author} {\bibfnamefont
  {S.}~\bibnamefont {Liuti}},\ }\href@noop {} {\  (\bibinfo {year} {2014})},\
  \Eprint {http://arxiv.org/abs/1401.0438} {arXiv:1401.0438 [hep-ph]}
  \BibitemShut {NoStop}%
\bibitem [{\citenamefont {Radici}\ \emph {et~al.}(2015)\citenamefont {Radici},
  \citenamefont {Courtoy}, \citenamefont {Bacchetta},\ and\ \citenamefont
  {Guagnelli}}]{Radici:2015mwa}%
  \BibitemOpen
  \bibfield  {author} {\bibinfo {author} {\bibfnamefont {M.}~\bibnamefont
  {Radici}}, \bibinfo {author} {\bibfnamefont {A.}~\bibnamefont {Courtoy}},
  \bibinfo {author} {\bibfnamefont {A.}~\bibnamefont {Bacchetta}}, \ and\
  \bibinfo {author} {\bibfnamefont {M.}~\bibnamefont {Guagnelli}},\ }\href
  {\doibase 10.1007/JHEP05(2015)123} {\bibfield  {journal} {\bibinfo  {journal}
  {JHEP}\ }\textbf {\bibinfo {volume} {05}},\ \bibinfo {pages} {123} (\bibinfo
  {year} {2015})},\ \Eprint {http://arxiv.org/abs/1503.03495} {arXiv:1503.03495
  [hep-ph]} \BibitemShut {NoStop}%
\bibitem [{\citenamefont {Pitschmann}\ \emph {et~al.}(2015)\citenamefont
  {Pitschmann}, \citenamefont {Seng}, \citenamefont {Roberts},\ and\
  \citenamefont {Schmidt}}]{Pitschmann:2014jxa}%
  \BibitemOpen
  \bibfield  {author} {\bibinfo {author} {\bibfnamefont {M.}~\bibnamefont
  {Pitschmann}}, \bibinfo {author} {\bibfnamefont {C.-Y.}\ \bibnamefont
  {Seng}}, \bibinfo {author} {\bibfnamefont {C.~D.}\ \bibnamefont {Roberts}}, \
  and\ \bibinfo {author} {\bibfnamefont {S.~M.}\ \bibnamefont {Schmidt}},\
  }\href {\doibase 10.1103/PhysRevD.91.074004} {\bibfield  {journal} {\bibinfo
  {journal} {Phys. Rev.}\ }\textbf {\bibinfo {volume} {D91}},\ \bibinfo {pages}
  {074004} (\bibinfo {year} {2015})},\ \Eprint {http://arxiv.org/abs/1411.2052}
  {arXiv:1411.2052 [nucl-th]} \BibitemShut {NoStop}%
\bibitem [{\citenamefont {Chen}\ \emph {et~al.}(2014)\citenamefont {Chen},
  \citenamefont {Gao}, \citenamefont {Hemmick}, \citenamefont {Meziani},\ and\
  \citenamefont {Souder}}]{Chen:2014psa}%
  \BibitemOpen
  \bibfield  {author} {\bibinfo {author} {\bibfnamefont {J.}~\bibnamefont
  {Chen}}, \bibinfo {author} {\bibfnamefont {H.}~\bibnamefont {Gao}}, \bibinfo
  {author} {\bibfnamefont {T.}~\bibnamefont {Hemmick}}, \bibinfo {author}
  {\bibfnamefont {Z.~E.}\ \bibnamefont {Meziani}}, \ and\ \bibinfo {author}
  {\bibfnamefont {P.}~\bibnamefont {Souder}} (\bibinfo {collaboration}
  {SoLID}),\ }\href@noop {} {\  (\bibinfo {year} {2014})},\ \Eprint
  {http://arxiv.org/abs/1409.7741} {arXiv:1409.7741 [nucl-ex]} \BibitemShut
  {NoStop}%
\end{thebibliography}
\end{document}